\documentclass[fleqn,usenatbib]{mnras}

\usepackage[T1]{fontenc}
\usepackage{ae,aecompl}

\usepackage{graphicx}	
\usepackage{amsmath}	
\usepackage{amssymb}	
\usepackage{subcaption}
\usepackage{gensymb}
\usepackage{tablefootnote}
\usepackage{footnote}
\makesavenoteenv{tabular}
\usepackage{tabularx}
\usepackage{float}

\title[Light curve properties of SN 2017fgc and HV SNe Ia]{Light curve properties of SN 2017fgc and HV SNe Ia}

\author[]{Umut Burgaz$^{1,2}$\thanks{E-mail:burgaz.umut@gmail.com},
    Keiichi Maeda$^{2}$,
    Belinda Kalomeni$^{1}$,
    Miho Kawabata$^{2}$,
    \newauthor
    Masayuki Yamanaka$^{3}$,
    Koji S. Kawabata$^{4,5,6}$,
    Naoki Kawahara$^{4,5}$
    \newauthor
    and Tatsuya Nakaoka$^{4,5}$
\\
$^{1}$Department of Astronomy and Space Sciences, University of Ege, 35100, \.{I}zmir, Turkey\\
$^{2}$Department of Astronomy, Kyoto University, Kitashirakawa-Oiwakecho, Sakyo-ku, Kyoto 606-8502, Japan\\
$^{3}$Okayama Observatory, Kyoto University, 3037-5 Honjo, Kamogata-cho, Asakuchi, Okayama 719-0232, Japan\\
$^{4}$Hiroshima Astrophysical Science Center, Hiroshima University, Kagamiyama 1-3-1, Higashi-Hiroshima ,Hiroshima 739-8526\\
$^{5}$Department of Physical Science, Hiroshima University, Kagamiyama 1-3-1, Higashi-Hiroshima 739-8526, Japan\\
$^{6}$Core Research for Energetic Universe (CORE-U), Hiroshima University, Kagamiyama, Higashi-Hiroshima, Hiroshima 739-8526, Japan
}

\date{Accepted 22 January 2021. Received YYY; in original form ZZZ}

\pubyear{2021}

\begin{document}
\label{firstpage}
\pagerange{\pageref{firstpage}--\pageref{lastpage}}
\maketitle

\begin{abstract}
Photometric and spectroscopic observations of type Ia supernova (SN) 2017fgc which cover the period from $-$12 to +137 days since the $B$-band maximum are presented. SN 2017fgc is a photometrically normal SN Ia with the luminosity decline rate, $ \Delta m _{15} (B)_{true} $= 1.10 $ \pm $ 0.10 mag. Spectroscopically, it belongs to the High Velocity (HV) SNe Ia group, with the Si II $\lambda$6355 velocity near the $B$-band maximum estimated to be 15,200  $ \pm $ 480 km $s^{-1}$. At the epochs around the near-infrared secondary peak,  the $R$ and $I$ bands show an excess of $\sim$0.2 mag level compared to the light curves of the normal velocity (NV) SNe Ia. Further inspection of the samples of HV and NV SNe Ia indicates that the excess is a generic feature among HV SNe Ia, different from NV SNe Ia. There is also a hint that the excess is seen in the V band, both in SN 2017fgc and other HV SNe Ia, which behaves like a less prominent shoulder in the light curve. The excess is not obvious in the B band (and unknown in the U band), and the color is consistent with the fiducial SN color. This might indicate the excess is attributed to the bolometric luminosity, not in the color. This excess is less likely caused by external effects, like an echo or change in reddening but could be due to an ionization effect, which reflects an intrinsic, either distinct or continuous, difference in the ejecta properties between HV and NV SNe Ia.

\end{abstract}




\section{Introduction}

Over the last few decades, type Ia supernovae (SNe Ia) have been important targets for transient observations since they have proven to be a reliable distance estimator. Thanks to their huge brightness, these SNe can be seen up to large distances, allowing measurements of fundamental cosmological parameters \citetext{\citealp{1998AJ....116.1009R}; \citealp{1999ApJ...517..565P}; \citealp{2016ApJ...826...56R}}.

Even though SNe Ia are acknowledged to be a thermonuclear explosion of a C/O white dwarf (WD) in a binary system \citep{1960ApJ...132..565H}, the explicit natures of the thermonuclear explosion and the binary companion are still under debate \citep{2016IJMPD..2530024M}. The commonly accepted progenitor models are divided into two scenarios. The first model is the single degenerate (SD) model \citep{1997Sci...276.1378N} in which the companion is a non-degenerate main-sequence star or an evolved star such as a red giant or a helium star. The second model is the double degenerate (DD) model \citep{1984ApJS...54..335I}, which involves the merger of two WDs. Different progenitor scenarios may manifest themselves in observational properties of SNe Ia. It is, therefore, crucial to understand similarities and diversities seen in SNe Ia.

The similarity among SNe Ia is highlighted by their photometric properties. It has been found that the absolute magnitudes of SNe Ia follow the width-luminosity relation \citetext{\citealp{1993ApJ...413L.105P}; \citealp{1996ApJ...473...88R}; \citealp{2005A&A...443..781G}; \citealp{2007A&A...466...11G}} and the color-luminosity relation \citetext{\citealp{1998A&A...331..815T}; \citealp{1999ApJ...525..209T}; \citealp{2005ApJ...620L..87W}}. On the other hand, their spectral properties point to a diversity among SNe Ia. \citet{2006PASP..118..560B} divided SNe Ia into four groups based on the empirical pseudo equivalent width (pEW) measurements of 5750{\AA} and 6100{\AA} absorption features: ``Core-Normal (CN)", "Broad Line (BL)", ``Cool (CL)", and ``Shallow-Silicon (SS)". \citet{2005ApJ...623.1011B} introduced a different classification scheme, based on the Si II $\lambda$6355 velocity evolution: The ``faint" SN 1991bg like group, the ``low velocity gradient (LVG)" group, and the ``high velocity gradient (HVG)" group. In general, HVG SNe Ia have a higher velocity than the faint and LVG SNe Ia. \citet{2009ApJ...699L.139W} proposed a similar classification based on the Si II $\lambda$6355 expansion velocity near maximum light ($v_{0}$); ``High Velocity (HV)", ($v_{0} \gtrsim$ 12,000 kms$^{-1}$) and ``Normal Velocity (NV)" ($v_{0} \lesssim$ 12,000 kms$^{-1}$) groups.

Aside from the line velocity evolution, HV SNe Ia exhibit other properties that are different from those of NV SNe Ia; HV SNe Ia tend to have a redder intrinsic $B-V$ color \citetext{\citealp{2008MNRAS.388..971P}; \citealp{2009ApJ...699L.139W}}, and they exhibit different evolution from NV SNe Ia in the $B$-band light curves \citetext{\citealp{2008ApJ...675..626W}; \citealp{2019ApJ...882..120W}} and the $B-V$ color at $\sim$40d after the maximum brightness \citep{2019ApJ...882..120W}. Different extinction laws seem to apply to the two types \citep{2009ApJ...699L.139W} where the extinction parameter, $R_{V} = A_{V} /E(B-V)$, is lower for HV SNe Ia ($R_{V} \sim$ 1.55) than for NV SNe Ia ($R_{V} \sim$ 2.51). It has also been shown that in the late epochs, the emission lines shift differently between HV and NV SNe Ia \citep{2010Natur.466...82M}. The explosions sites of HV SNe Ia tend to be more concentrated near the core of the host galaxies than those of NV SNe Ia \citep{2013Sci...340..170W}, which might indicate different progenitor systems. The origin of the differences between the two sub-types of SNe Ia is still not well understood, and more observational and theoretical studies are needed. 

In this paper, we present the observations of SN 2017fgc, a HV SN Ia that was discovered by DLT40 \citep{{2017TNSCR.757....1V}} on 2017 July 11. T\"{U}B\.{I}TAK National Observatory (TUG) picked up the alert on 2017 July 13 (MJD 57947.011) with the TUG T60 telescope, and started observing the SN around 11 days before the $B$-band maximum (MJD 57958.932). Based on a spectrum obtained by \citet{2017ATel10569....1S}, the object was classified as an SN Ia around $-$13d before the maximum with a red-shift of z=0.008. Our observations and data reduction of SN 2017fgc are presented in Section~\ref{sec:obs}. Light curve properties, reddening estimate, absolute magnitudes, and the optical spectra of SN 2017fgc are shown in Section~\ref{sec:prop2017fgc}. Further comparison of $BVRI$ light curves of NV and HV SNe Ia is presented in Section~\ref{sec:LCcomp}, where we find differences in the light curve evolution between HV SNe Ia (highlighted by SN 2017fgc) and NV SNe Ia. We explore this property further in Section~\ref{sec:excess}. Conclusions are presented in Section~\ref{sec:conc}.

\section{Observations and data reduction}
\label{sec:obs}

\begin{figure}
   \centering
     \includegraphics[width=\columnwidth]{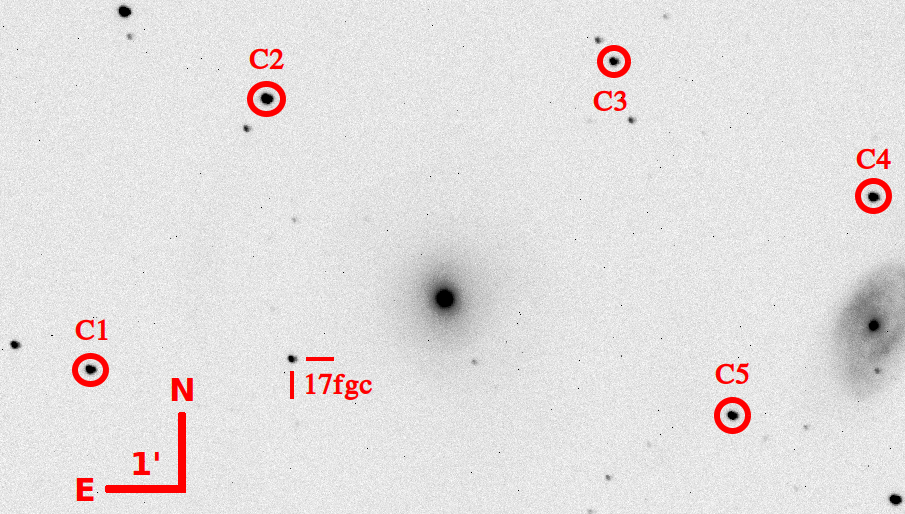}
   \caption{The $V$-band image of SN 2017fgc taken by TUG T60 telescope on 2017 July 25 with local comparison stars being also marked.}
   \label{fig:T60ImageComp}
 \end{figure}

SN 2017fgc was discovered by DLT40 on 2017 July 11 (MJD 57945.469) and the alert was published at Transient Name Server\footnote{\label{TNS}https://wis-tns.weizmann.ac.il/} on the same day (MJD 57945.472) when it was at 17.3 mag (unfiltered). The last non-detection was on 2017 July 8 (MJD 57942.291) with a limiting magnitude of 19.5 mag. The coordinate of SN 2017fgc is $\alpha =01^h20^m14^s.44$ and $\delta$ = $03\degree24'09''.96$ (J2000.0). SN 2017fgc is located on $116''.0$ east and $45''.0$ north of the centre of NGC 0474. This is a famous shell galaxy with tidal tails \citetext{\citealp{1983ApJ...274..534M}; \citealp{1999MNRAS.307..967T}} and is in interaction with NGC 0470 by a HI gas bridge \citep{2006MNRAS.368..851R}, and both of these two galaxies lie at a distance of $\sim$ 27.7 Mpc \citep{2012ApJ...753...43K}. A TUG T60 image of SN 2017fgc taken in V band is presented in Fig.~\ref{fig:T60ImageComp}.

\subsection{Imaging}

Imaging observations were initiated soon after the alert with the 0.6 m robotic TUG T60 telescope at T\"{U}B\.{I}TAK National Observatory, on 2017 Jul 13 (JD 2457947.511340) and conducted until 2018 January 29. The observations have been performed by using the FLI ProLine PL3041 CCD camera ($2048 \times 2048$ pixels, covering a $17.4 \times 17.4$ $arcmin^2$ field of view) using the $BVRI$ photometric passbands. In total, the data are taken on 65 nights starting from $\sim$12 days before the $B$-band maximum to $\sim$137 days after.

Bias subtraction, dark, and flat-field corrections were done by using the CCDproc package under Astropy. The photometric reduction was done by giving the WCS coordinates to the stars using Astrometry.net\footnote{\label{astrometry}nova.astrometry.net} and obtaining a catalog file with the instrumental magnitudes by running SExtractor \citep{{1996A&AS..117..393B}} on the images. Considering the distance from the SN to the host galaxy, no galaxy subtraction was done on the images. The preliminary photometric calibration and conversion from instrumental magnitudes to apparent magnitudes using differential photometry was carried out by the Cambridge Photometry Calibration Server (CPCS)\footnote{\label{CPCS}http://gsaweb.ast.cam.ac.uk/followup/}, which made the data available to all observers during the whole observing periods. Final photometric calibration was done using calibration stars from the Data Release 10 \citep{2018AAS...23222306H} of The AAVSO Photometric All-Sky Survey (APASS). The $B$-band and $V$-band data were taken from the original APASS values directly. The $R$-band and the $I$-band magnitudes of the calibrating stars were obtained by converting APASS photometry of the $r$ and $i$ bands to the Johnson $R$ and $I$ bands using the following SDSS photometry transformation equations; $R = r - 0.2936 \times (r - i) - 0.1439 \pm 0.007$ mag and $I = r - 1.2444 \times (r - i) - 0.3820 \pm 0.008$ mag. The zero-points with an average error of 0.02 mag were obtained for each data individually. First-order color term correction was applied in the photometry. The apparent magnitudes of SN 2017fgc are listed in Table~\ref{tab:tugdata}.

\subsection{Spectroscopy}

A total of 10 spectra of SN 2017fgc were obtained with the Hiroshima One$-$shot Wide-field Polarimeter \citetext{HOWPol; \citealp{2008SPIE.7014E..4LK}} installed on the Nasmyth stage of the 1.5$-$m Kanata telescope at the Higashi-Hiroshima Observatory, Hiroshima University. HOWPol has  wavelength coverage of 4500$-$9200\r{A} and R = $\lambda$/$\Delta\lambda$ $\simeq$ 400 at 6000\r{A}. Sky emission lines were used for the wavelength calibration. L. A. Cosmic pipeline \citep{2001PASP..113.1420V} was used to remove cosmic ray events. Flux calibration was done by using the data of spectrophotometric standard stars observed on the same night. The log of spectroscopy is presented in Table~\ref{tab:speclog}. The identification spectrum \citep{2017ATel10569....1S} from the DLT survey at $-$13 days is also included in our analysis.

\begin{table}
	\centering
\begin{minipage}[b]{\columnwidth}
	\caption{A log of the spectroscopic observations of SN 2017fgc.}
	\label{tab:speclog}
	\begin{tabular}{ccccc} 
		\hline
		Date & MJD & Phase$^{a}$ & Coverage & Resolution \\
		 & & (days) & (\r{A}) & (\r{A})\\
		\hline
		13/07/17 & 57947.8 & $-$10.9 & 4500$-$9200 & 500\\
		19/07/17 & 57953.8 & $-$4.9 & 4500$-$9200 & 500\\
		30/07/17 & 57964.7 & 6.0 & 4500$-$9200 & 500\\
		31/07/17 & 57965.7 & 7.0 & 4500$-$9200 & 500\\
		01/08/17 & 57966.7 & 8.0 & 4500$-$9200 & 500\\
		03/08/17 & 57968.7 & 10.0 & 4500$-$9200 & 500\\
		20/08/17 & 57985.7 & 27.0 & 4500$-$9200 & 500\\
		27/08/17 & 57992.7 & 34.0 & 4500$-$9200 & 500\\
		01/09/17 & 57997.7 & 39.0 & 4500$-$9200 & 500\\
		25/09/17 & 58021.6 & 62.9 & 4500$-$9200 & 500\\		
		\hline
	\end{tabular}
	       {\raggedright} ($^a$)Relative to the $B$-band maximum (MJD = 57958.7)
	     \end{minipage}
\end{table}

\section{Properties of SN 2017fgc}
\label{sec:prop2017fgc}

\subsection{Optical Spectra}
\label{sec:optspec}
\begin{figure}
	 \centering
	 \includegraphics[width=\columnwidth]{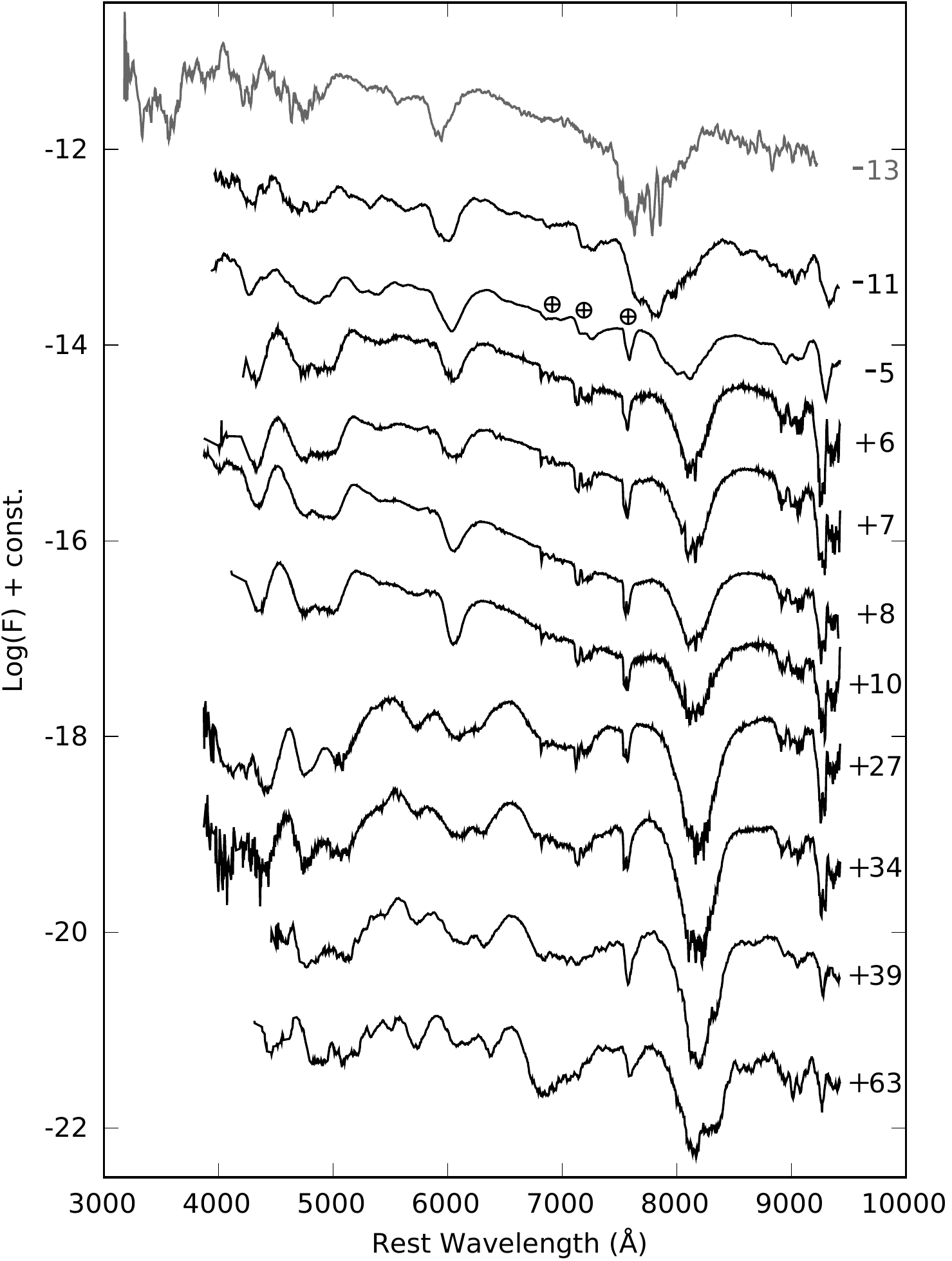}
     \caption{Optical spectral evolution of SN 2017fgc. The spectra have been corrected for the redshift of NGC 474 ($v_{hel}$ = 2315 kms$^{-1}$). No reddening correction is applied. The spectra have been artificially shifted in the vertical axis for clarity. The numbers on the right side show the epochs of the spectra in days after the $B$-band maximum.}
     \label{fig:specevo}
	 \end{figure}

\begin{figure}
	 \centering
	 \includegraphics[width=\columnwidth]{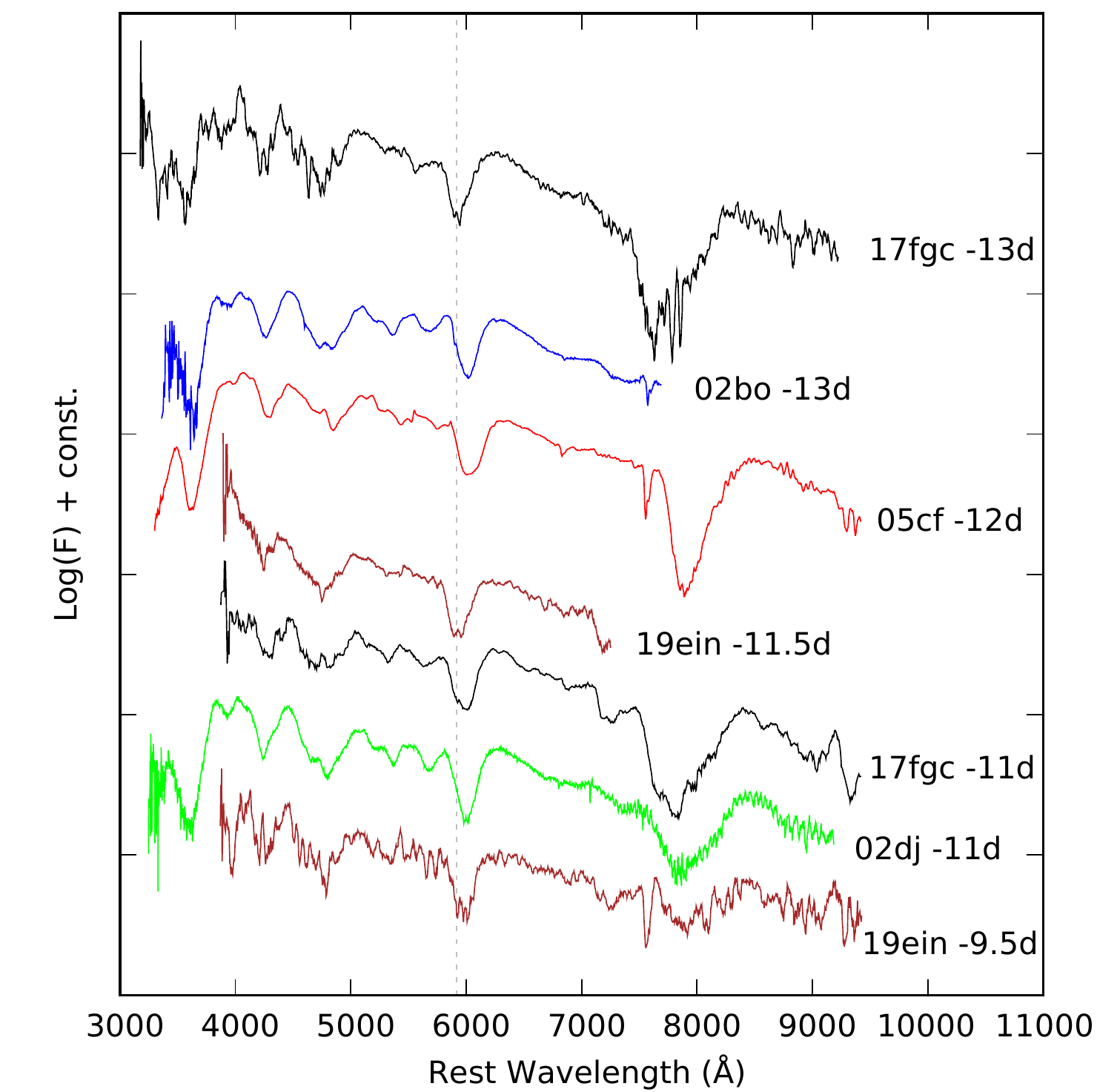}
     \caption{Spectrum of SN 2017fgc at t = $-$13 and t = $-$11 days since the $B$-band maximum, over-plotted with the spectra of SN 2019ein, SN2002dj, SN 2002bo and SN 2005cf around similar phases (see the text for the references). All the presented spectra have been corrected for redshift and reddening. The dashed line marks the absorption minima of Si II $\lambda$6355 for SN 2017fgc at t = $-$13d.}
     \label{fig:minus13evo}
	 \end{figure}

The spectral series cover the phase from t $\simeq$ $-$11 to t $\simeq$ +63 days since the $B$-band maximum. The complete spectral evolution of SN 2017fgc, including the identification spectra from DLT (shown with gray color), is presented in Fig.~\ref{fig:specevo}. The earliest spectra show a strong absorption feature at $\sim$5900\AA, by Si II $\lambda$6355, indicating a high velocity. Although the spectral inspection of SN 2017fgc is not the main focus of this paper, comparisons in the earliest phase with SN 2019ein \citep{2020ApJ...893..143K}, SN 2002dj \citep{2008MNRAS.388..971P}, SN 2002bo \citep{2004MNRAS.348..261B} and SN 2005cf \citep{2007A&A...471..527G} are presented in Fig. \ref{fig:minus13evo}. Comparisons in other phases are presented in Fig. \ref{fig:SpecComp}. All the spectra have been corrected for redshifts and reddening (see Section~\ref{sec:optspec}), except for those in Fig. \ref{fig:specevo}.

\begin{figure}
	 \centering
	 \includegraphics[width=\columnwidth]{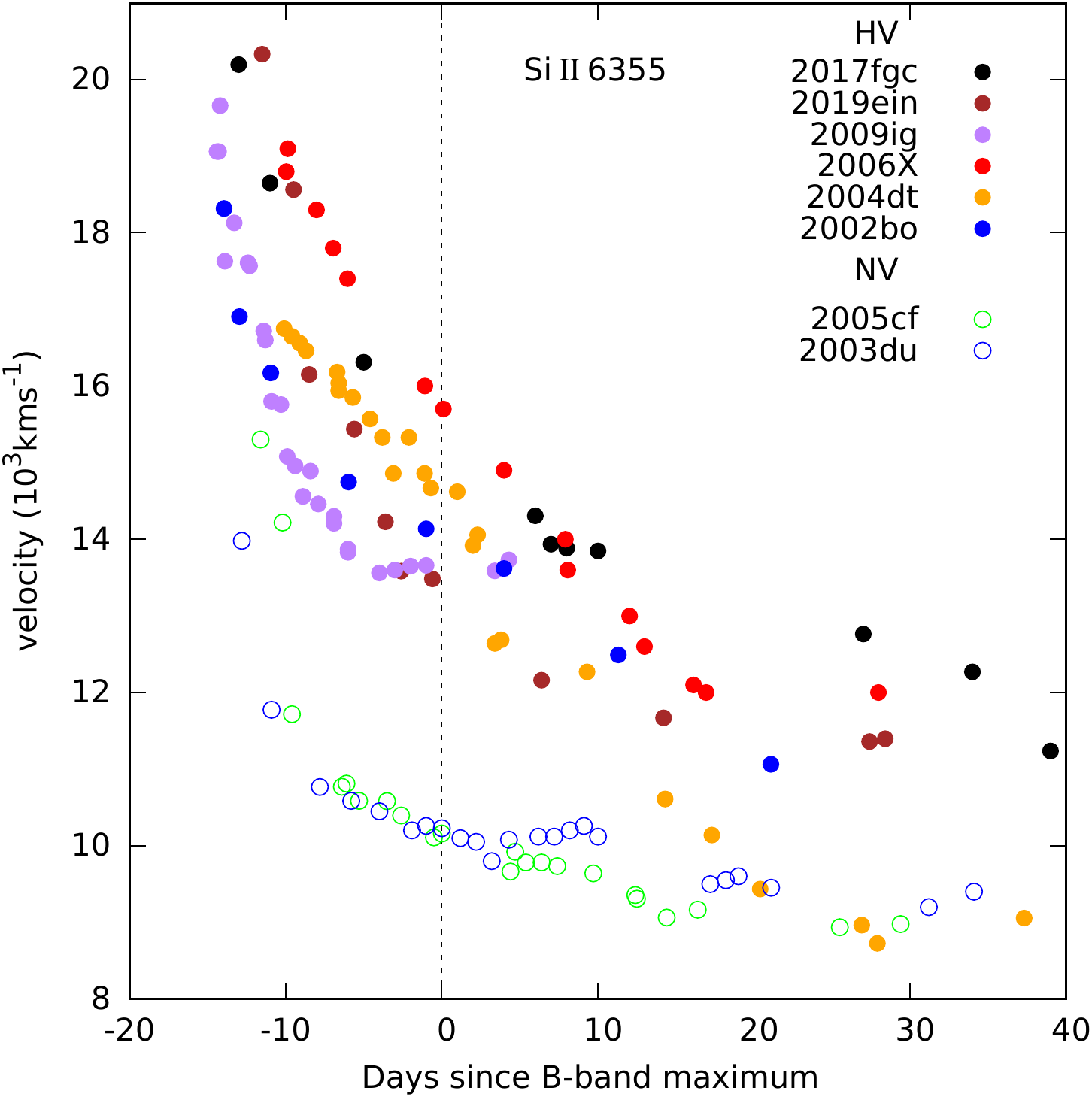}
     \caption{Expansion velocity evolution of SN 2017fgc as measured from the minima of Si II $\lambda$6355, along with SN 2019ein \citep{2020ApJ...893..143K}, SN 2009ig \citep{2015MNRAS.451.1973S}, SN 2006X \citep{2009PASJ...61..713Y}, SN 2004dt \citep{2007A&A...475..585A}, SN 2002bo \citep{2004MNRAS.348..261B}, SN 2005cf \citep{2007A&A...471..527G} and SN 2003du \citep{2007A&A...469..645S}.}
     \label{fig:Speed}
	 \end{figure}

By following the method described in \citet{2015ApJS..220...20Z} and performing a single Gaussian fitting to the observed line profile of Si II $\lambda$6355, the expansion velocity ($v_{exp}$) of SN 2017fgc is derived and presented in Fig.~\ref{fig:Speed} as the function of time, together with those of other well-observed SNe Ia. All velocities have been corrected for the redshift. It can be seen that SN 2017fgc has a very high expansion velocity with $v_{exp}$ reaching to $\sim$ 20,500 km $s^{-1}$ at t = $-$13 d. Near the $B$-band maximum, from a polynomial fit, $v_{exp}$ is estimated to be 15,200  $ \pm $ 480 km $s^{-1}$ while the average value for NV SNe Ia is around 11,800 km $s^{-1}$ at maximum light \citep{2009ApJ...699L.139W}. Therefore, SN 2017fgc is classified into the HV SN Ia subclass.

\subsection{BVRI light curves}

\begin{figure}
	 \centering
	 \includegraphics[width=\columnwidth]{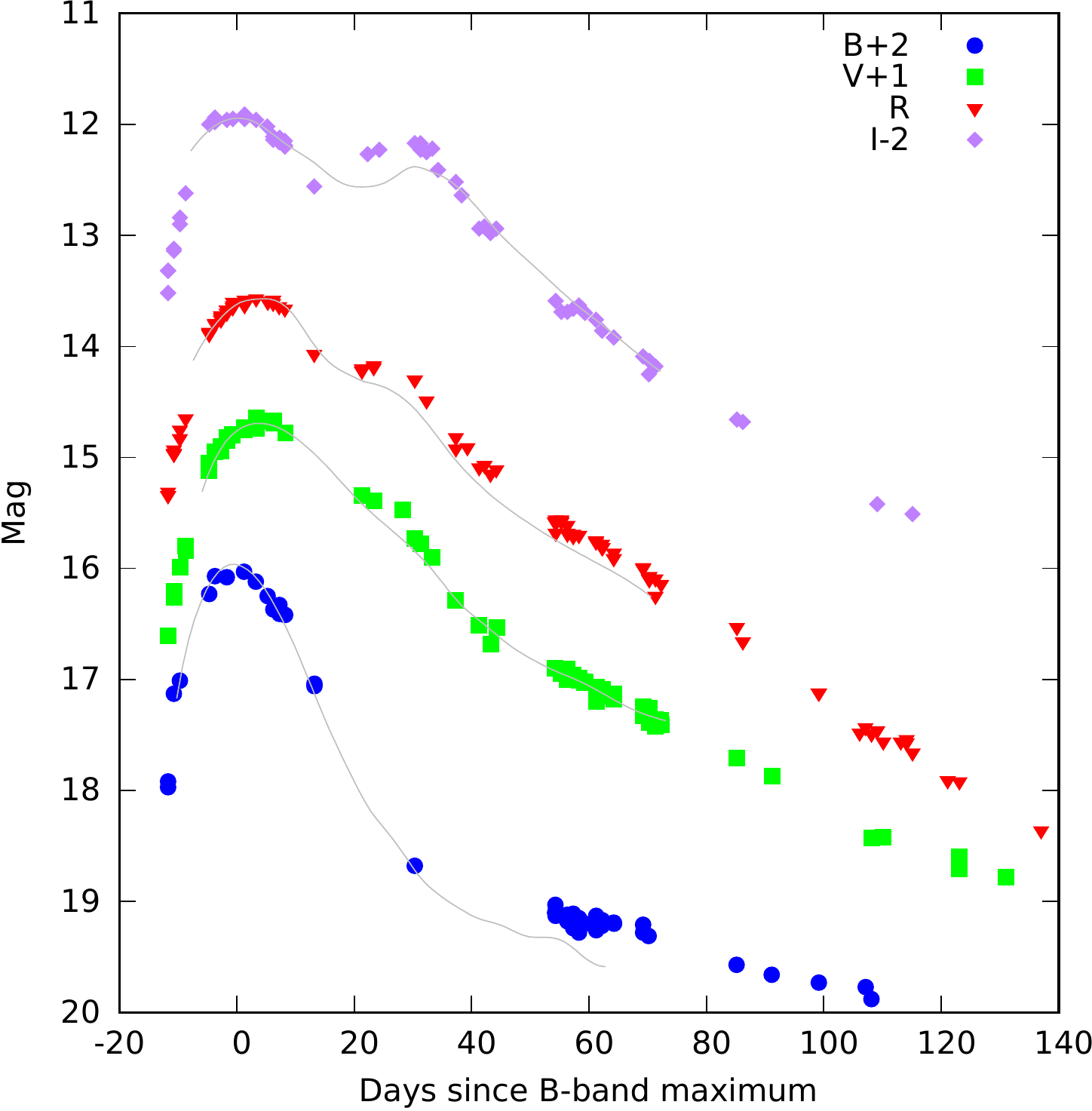}
     \caption{$BVRI$ light curves of SN 2017fgc. The gray line represents the best SNooPy fit in each band.}
     \label{fig:SN 2017fgc_bvri}
	 \end{figure}

The $BVRI$ light curves of SN 2017fgc are presented in Fig.~\ref{fig:SN 2017fgc_bvri}. Thanks to the good photometric coverage, the maximum epoch is well constrained. A polynomial fit is applied to the points around the maximum, and $B_{max}$ = 13.97 $\pm$ 0.05 mag on $t_{Bmax}$ = MJD 57958.6 $\pm$ 0.8 was obtained. Thus, the observations cover the epochs between $-$12 and +137 days with respect to $t_{Bmax}$. The maximum magnitudes in the other bands are the following; 13.64 $\pm$ 0.03 mag ($V$), 13.57 $\pm$ 0.01 mag ($R$), and 13.91 $\pm$ 0.01 mag ($I$) on $\sim$3, 2, and 0.4 days after the $B$-band maximum, respectively. 

By using the light curve fitter SNooPy \citep{2011AJ....141...19B}, basic light curve parameters were computed. Both the ``EBV model" and ``max model" in SNooPy were used for the analysis. These models fit the $BVRI$ light curves with the templates generated with the prescription given by \citet{2006ApJ...647..501P}. The process of the light curve fitting starts with an initial fit to determine the time of $t_{Bmax}$, which allows for the initial K-corrections to be determined by applying the spectral energy distribution (SED) templates from \citet{2007ApJ...663.1187H}. Afterward, the K-corrected data go through another fit, which allows colors to be computed as a function of time. Then, by warping the SED to match to the observed colors, an improved K-correction is computed. Finally, the last fit is performed using the newly calculated improved K-corrections. In both ``EBV model" and ``max model", SNooPy corrects the data for the MW extinction \citep{2011ApJ...737..103S}.

From the SNooPy analysis, $t_{Bmax}$ is found to have occurred on the MJD = 57958.721 $\pm$ 0.693. The decline rate, $\Delta m _{15} (B) $, with the K-correction is found to be 1.08 $ \pm $ 0.09 mag. It is consistent with a direct measurement from the light curve which gives $\Delta m _{15} (B) $ = 1.07 $ \pm $ 0.11 mag. The distance modulus and host galaxy reddening were also computed with SNooPy as well (see Section \ref{sec:RE}). 

Given that the direct measurement and the SNooPy fit give the consistent results, $t_{Bmax}$ = MJD 57958.721 $\pm$ 0.693 and $\Delta m _{15} (B) $ = 1.08 $ \pm $ 0.09 mag are adopted in this paper. Generally, $ \Delta m _{15} (B) $ for SNe Ia range from 0.72 \citep{2010AJ....139..120F} to 1.95 mag \citep{2010ApJS..190..418G}  with a typical value of 1.1 \citep{1999AJ....118.1766P}. Thus, SN 2017fgc is classified as a normal decliner. Secondary peaks are seen in the $R$ and $I$ band light curves, as expected for a normal decliner. The secondary peak in the $I$ band occurred around 30 days after the $B$-band maximum and was $\sim$ 0.25 mag fainter than the maximum brightness.

\subsection{Reddening Estimate and Absolute Magnitude}
\label{sec:RE}
The galactic reddening in the direction of SN 2017fgc is estimated to be $E(B-V)_{Gal}$ = 0.029 mag \citep{2011ApJ...737..103S}. The intrinsic $B-V$ color of SNe Ia in the maximum light can be estimated by several approaches. Using the relation between the B $-$ V color and $\Delta m _{15} (B)$ suggested by \citet{1999AJ....118.1766P}, $E(B$-$V)_{host}$ = 0.29 $\pm$ 0.02 mag is estimated, while using the approach suggested by \citet{2005ApJ...624..532R}, $E(B$-$V)_{host}$ = 0.25 $\pm$ 0.09 mag is estimated. Another approach with the relation between the Si II $\lambda$6355 line velocity and the intrinsic $B-V$ color of SNe Ia in the maximum light can also be used to estimate the host galaxy extinction. From \citet{2011ApJ...742...89F}, we derive $E(B-V)_{host}$ = 0.17 $\pm$ 0.07 mag, and by \citet{2012AJ....143..126B}, $E(B$-$V)_{host}$ = 0.29 $\pm$ 0.09 mag is estimated. The estimated host galaxy reddening obtained with the light curve fitter SNooPy is $E(B-V)_{host}$ = 0.36 $ \pm $ 0.11 mag with $R_{V}$=1.5. Since all the estimated values are in agreement, $E(B$-$V)_{host}$ = 0.29 $\pm$ 0.02 mag is adopted for the host galaxy reddening in this paper, leading to $E(B-V)_{tot}$ = 0.32 $\pm$ 0.02 mag.

An observed $\Delta m _{15} (B) $ = 1.08 $ \pm $ 0.09 mag is measured for SN 2017fgc. Applying the interstellar extinction suffered by SN 2017fgc, a correction to the $\Delta m _{15} (B)$ is applied by using the Equation 6 of \citet{1999AJ....118.1766P}. The reddening corrected decline rate parameter is then found to be $\Delta m _{15} (B)_{true} $ = 1.10 $\pm$ 0.10 mag.

The absolute magnitude of SN 2017fgc can be estimated by several methods. By using the relation between $M_B$ and $ \Delta m _{15} $ from \citet{1999AJ....118.1766P}, we derive M$_B$ = $-$19.38 $\pm$ 0.23 mag, which corresponds to a distance modulus of DM = 32.81 $\pm$ 0.33 with H$_{0}$ = 72 kms$^{-1}$Mpc$^{-1}$. By using the relation updated by \citet{2006ApJ...647..501P}, we found M$_B$ = $-$19.29 $\pm$ 0.14 mag, and DM = 32.72 $\pm$ 0.34 mag. By another method proposed by  \citet{2005ApJ...620L..87W}, which uses the relation between $M_B$ and the $B-V$ color $\sim$12 days after the $B$-band maximum ($\Delta C_{12}$), we obtain $M_B$ = $-$19.29 $ \pm $ 0.24 mag, and DM = 32.72 $\pm$ 0.31 mag. \citet{2004MNRAS.349.1344A} presented an updated version of \citet{1996AJ....112.2408H}, which gives M$_B$ = $-$19.40 $\pm$ 0.23 mag. DM obtained by all of the methods (extinction corrected, assuming R$_{V}$ = 3.1 for the Milky Way (MW), and R$_{V}$ = 1.55 for the host galaxy) agree with the distance to NGC 0474 within the errors. The results are listed in Table~\ref{tab:BmaxValues}.

\begin{table}
	\centering
\begin{minipage}[b]{\columnwidth}
	\caption{Estimated peak absolute magnitudes for SN 2017fgc.}
	\label{tab:BmaxValues}
	\begin{tabular}{cc}
		\hline
		Method & $M_{Bmax}$(mag)\\
		\hline
		Phillips et al. (1999) & $-$19.38 $ \pm $ 0.18 \\
		Altavilla et al. (2004) & $-$19.40 $ \pm $ 0.19 \\
		Wang et al. (2005) & $-$19.29 $ \pm $ 0.19 \\
		Prieto et al. (2006) & $-$19.29 $ \pm $ 0.14 \\
		\\
		Average & $-$19.34 $ \pm $ 0.19\\
		\hline
	\end{tabular}
	     \end{minipage}
\end{table}

\section{Light curve comparisons}
\label{sec:LCcomp}

\begin{figure}
	 \centering
	 \includegraphics[width=\columnwidth]{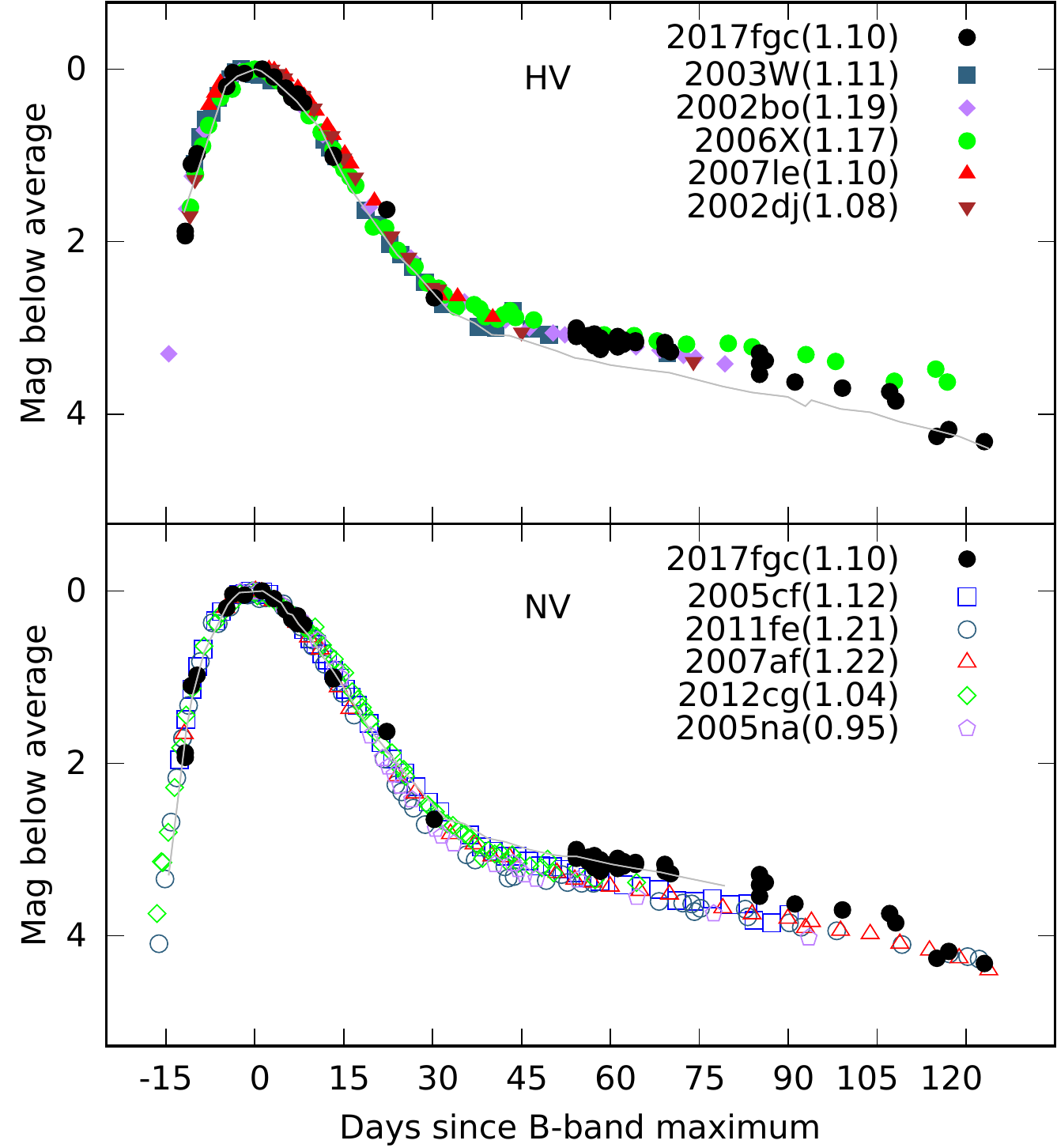}
     \caption{The $B$-band light curve of SN 2017fgc together with those of comparison SNe Ia. HV SNe Ia are plotted on the top, while NV SNe Ia are plotted on the bottom. All the light curves are shifted to match the time of $B$ maximum and the peak magnitude in the $B$ band. Gray lines represent NV SN 2007af (Top) and HV SN 2012bo (Bottom).}
     \label{fig:BBAND}
	 \end{figure}

\begin{figure}
	 \centering
	 \includegraphics[width=\columnwidth]{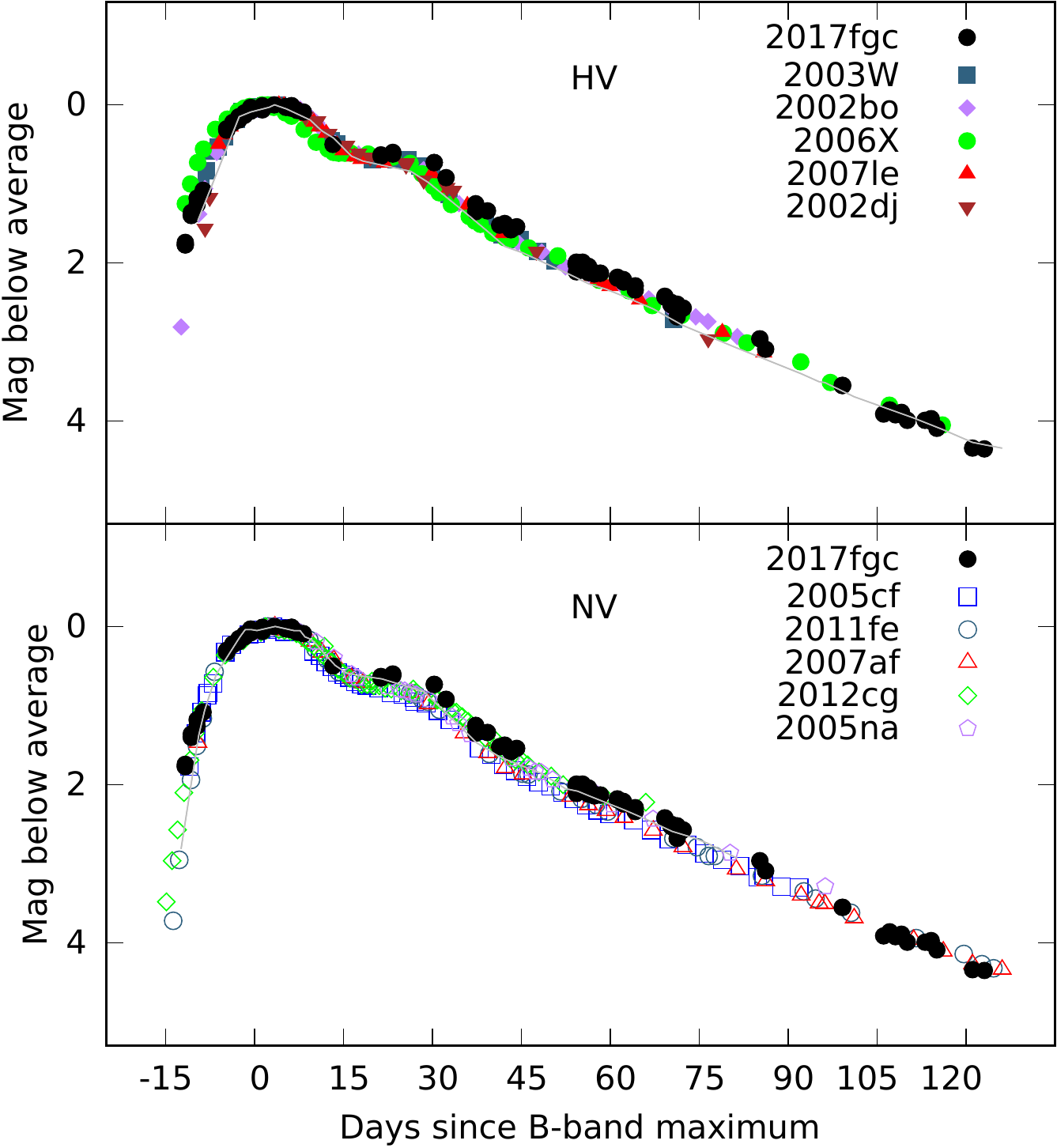}
     \caption{Same as Figure 6, but for the $R$ band.}
     \label{fig:RBAND}
	 \end{figure}

\begin{figure}
	 \centering
	 \includegraphics[width=\columnwidth]{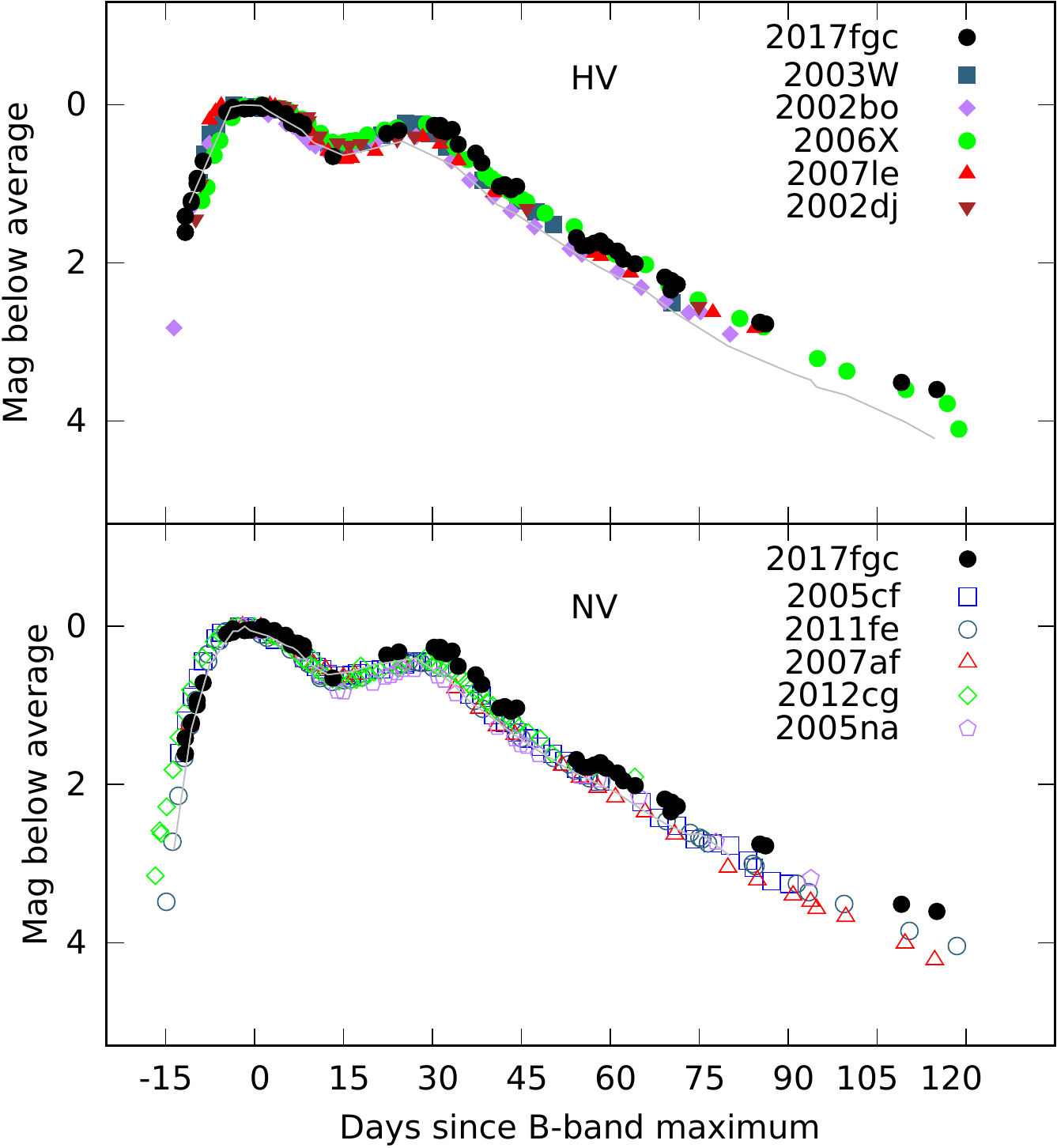}
     \caption{Same as Figure 6, but for the $I$ band.}
     \label{fig:IBAND}
	 \end{figure}

\begin{figure*}
   \centering
   \begin{subfigure}[b]{0.48\linewidth}
     \includegraphics[width=\linewidth]{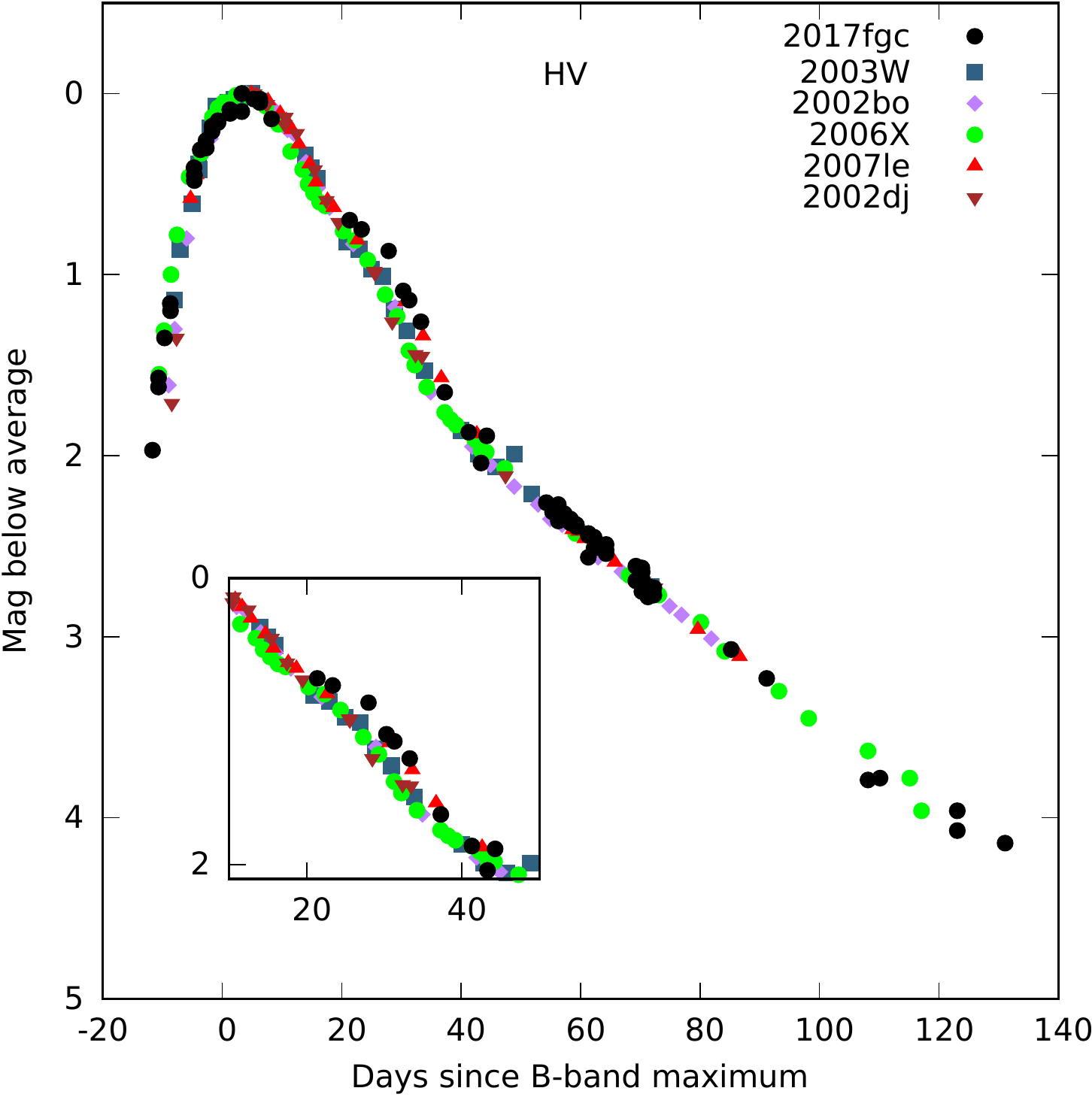}
   \end{subfigure}
   \begin{subfigure}[b]{0.48\linewidth}
     \includegraphics[width=\linewidth]{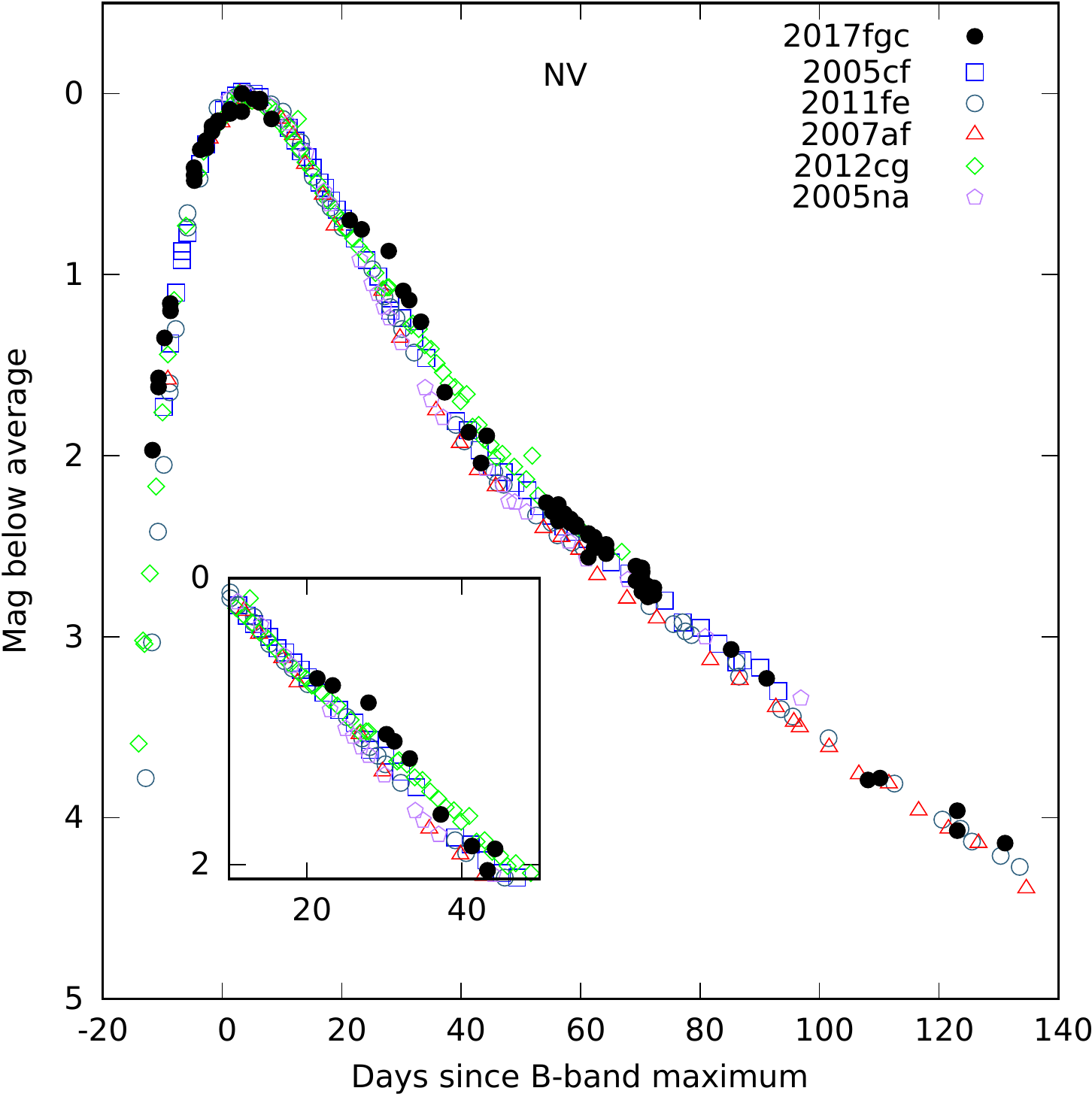}
   \end{subfigure}
   \caption{The $V$-band light curve of SN 2017fgc together with those of comparison SNe Ia. All the light curves are shifted to match the time of $B$ maximum and  peak magnitude in the $B$ band.}
   \label{fig:VBAND}
 \end{figure*}

We compare the multi-band light curves of SN 2017fgc to those of the following objects;  SN 2002dj [$ \Delta m _{15} (B) $ = 1.08; \citet{2008MNRAS.388..971P}], SN 2002bo [$ \Delta m _{15} (B) $ = 1.19; \citet{2010ApJS..190..418G}; \citet{2012MNRAS.425.1789S}], SN 2003W [$ \Delta m _{15} (B) $ = 1.11; \citet{2010ApJS..190..418G}; \citet{2012MNRAS.425.1789S}], SN 2006X [$ \Delta m _{15} (B) $ = 1.17; \citet{2008ApJ...675..626W}], SN 2005cf [$ \Delta m _{15} (B) $ = 1.12; \citet{2007MNRAS.376.1301P}], SN 2005na [$ \Delta m _{15} (B) $ = 0.95; \citet{2010AJ....139..519C}], SN 2007af [$ \Delta m _{15} (B) $ = 1.22; \citet{2010ApJS..190..418G}; \citet{2012MNRAS.425.1789S}], SN 2007le [$ \Delta m _{15} (B) $ = 1.10; \citet{2012ApJS..200...12H}], SN 2011fe [$ \Delta m _{15} (B) $ = 1.21; \citet{2012JAVSO..40..872R}; \citet{2016ApJ...820...67Z}], and SN 2012cg [$ \Delta m _{15} (B) $ = 1.04; \citet{2012MNRAS.425.1789S}; \citet{2013NewA...20...30M}]. All the light curves are normalized to the time of $B_{max}$ and to the peak magnitudes in the corresponding bands. Figs. \ref{fig:BBAND} $-$ \ref{fig:CC} show the comparisons. In each band, the comparison is done with a sample of HV SNe Ia and that of NV SNe Ia separately.

Fig.~\ref{fig:BBAND} reveals that the $B$-band light curve of SN 2017fgc is similar to both NV and HV SNe Ia until $\sim$ 50 days after the $B$-band maximum. The pre-maximum phase of SN 2017fgc in the $B$-band is most similar to SN 2002bo and SN 2003W. Around 50 days after the $t_{Bmax}$, SN 2017fgc starts to differ from those of NV SNe Ia, as previously shown by \citet{2019ApJ...882..120W} for HV SNe Ia.

We find that SN 2017fgc shows a difference from NV SNe Ia in the $V$ band with a shoulder-like behaviour that can be seen between $\sim$20 and $\sim$30 days after the $B$ band maximum (Fig.~\ref{fig:VBAND}). Similar behaviour was previously discovered in SN 2006X \citep{2008ApJ...675..626W}. Indeed, this behavior is commonly seen in other comparison HV SNe Ia (See Section \ref{sec:excess} for details).

SN 2017fgc differs from NV SNe Ia in the $R$-band light curve with a higher magnitude rise around the secondary peak as well (Fig.~\ref{fig:RBAND}). Indeed, it can be seen from Fig.~\ref{fig:RBAND} that HV SNe Ia generally seem to differ from NV SNe Ia in the $R$-band light curves. Similar behavior is also noticed in the $I$-band light curves (Fig.~\ref{fig:IBAND}). Note that these differences between HV and NV SNe Ia happen at the same time window in all the bands. The observed behavior of the secondary peak and the difference between HV and NV SNe Ia will be further discussed in Section \ref{sec:excess}. 

\subsection{Color curves}

\begin{figure*}
   \centering
   \begin{subfigure}[b]{0.49\linewidth}
     \includegraphics[width=\linewidth]{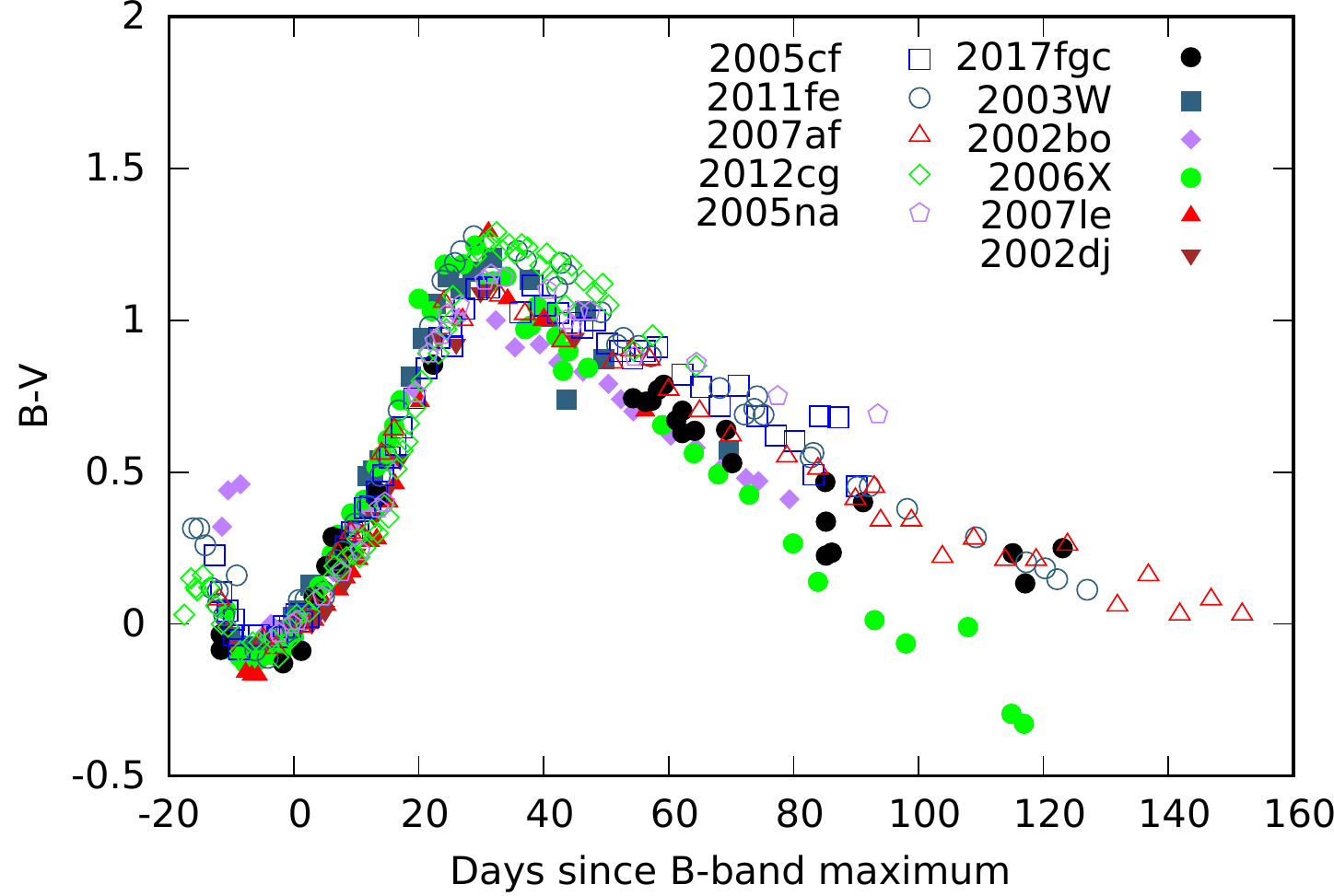}
   \end{subfigure}
   \begin{subfigure}[b]{0.49\linewidth}
     \includegraphics[width=\linewidth]{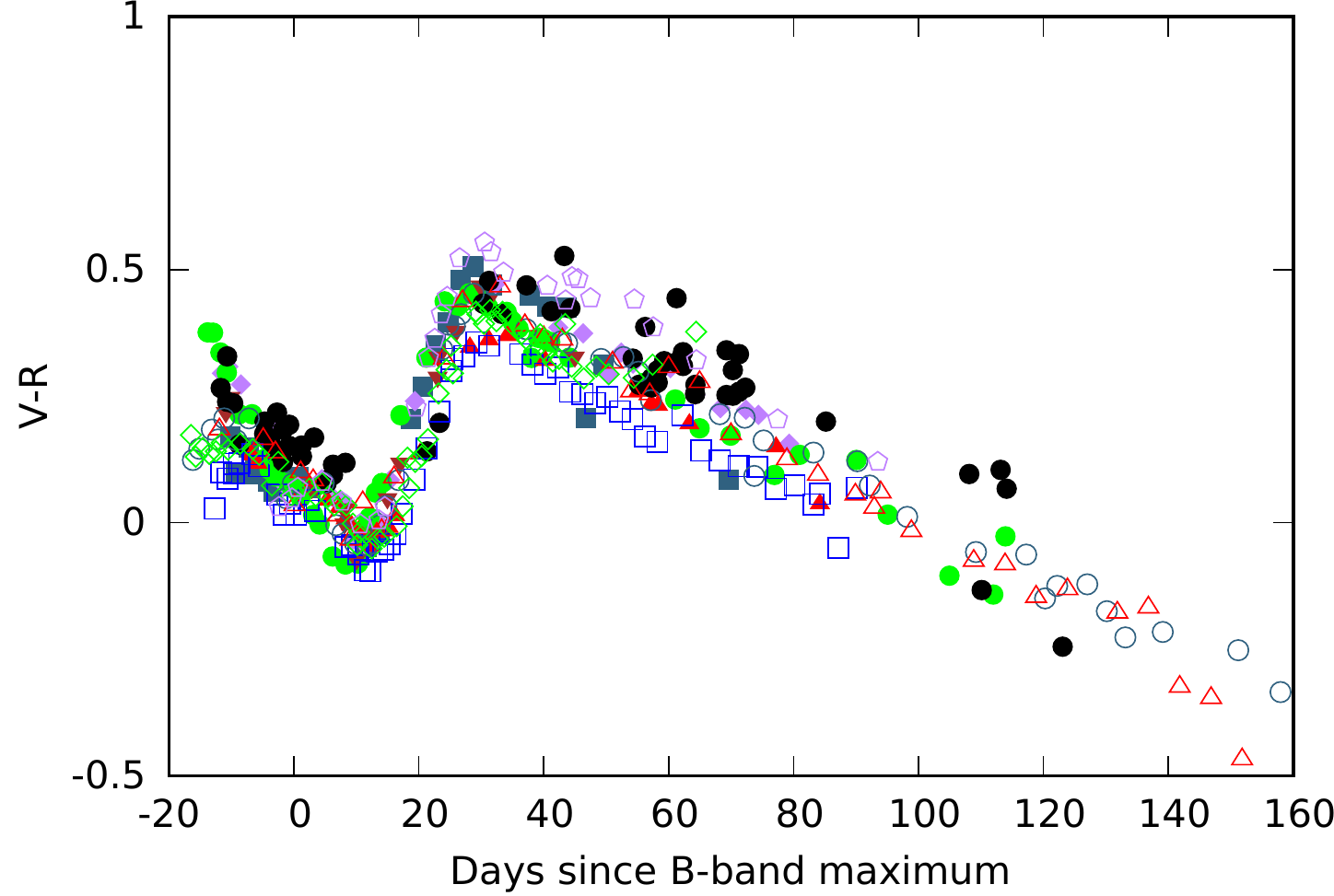}
   \end{subfigure}
   \begin{subfigure}[b]{0.49\linewidth}
     \includegraphics[width=\linewidth]{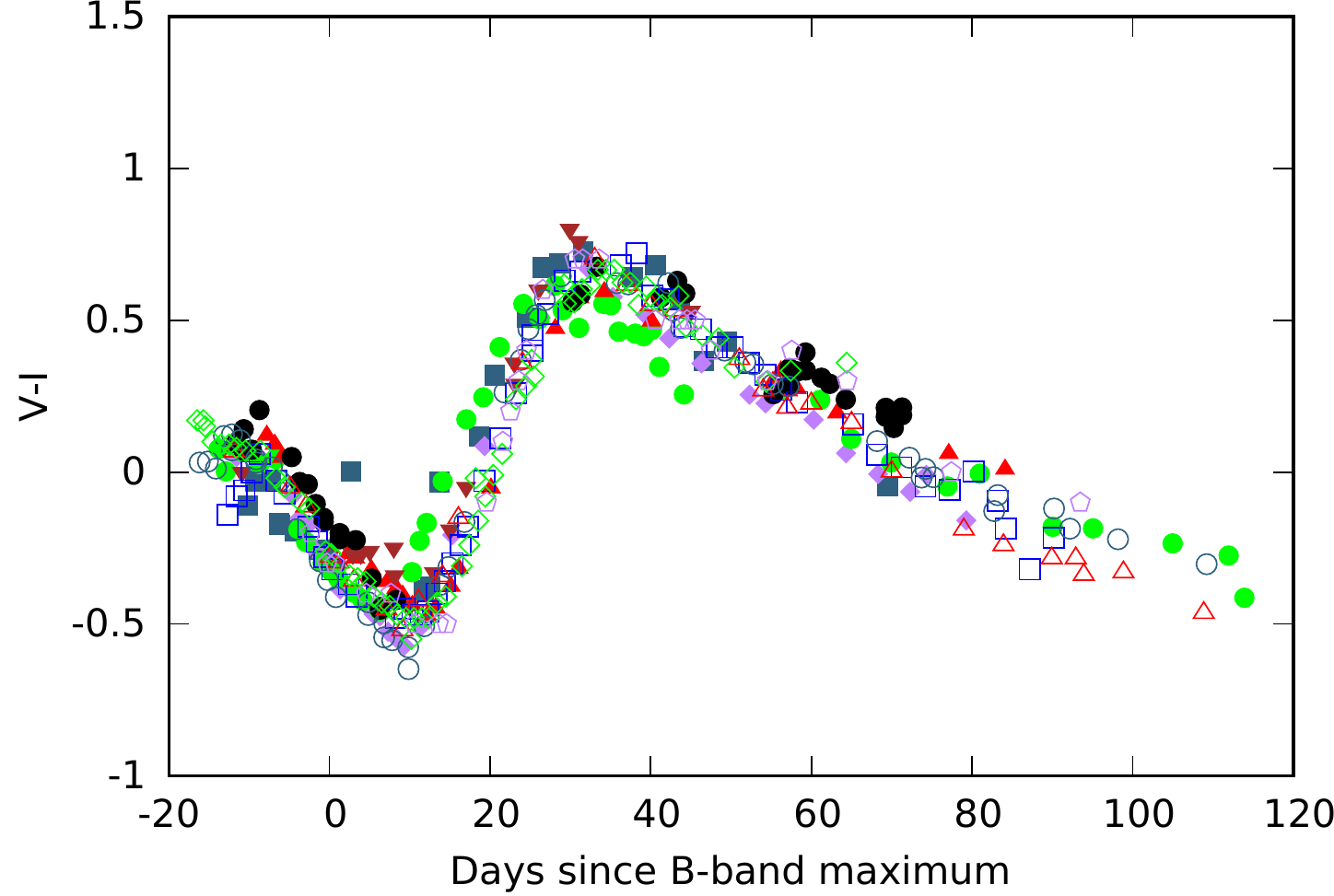}
   \end{subfigure}
   \begin{subfigure}[b]{0.49\linewidth}
     \includegraphics[width=\linewidth]{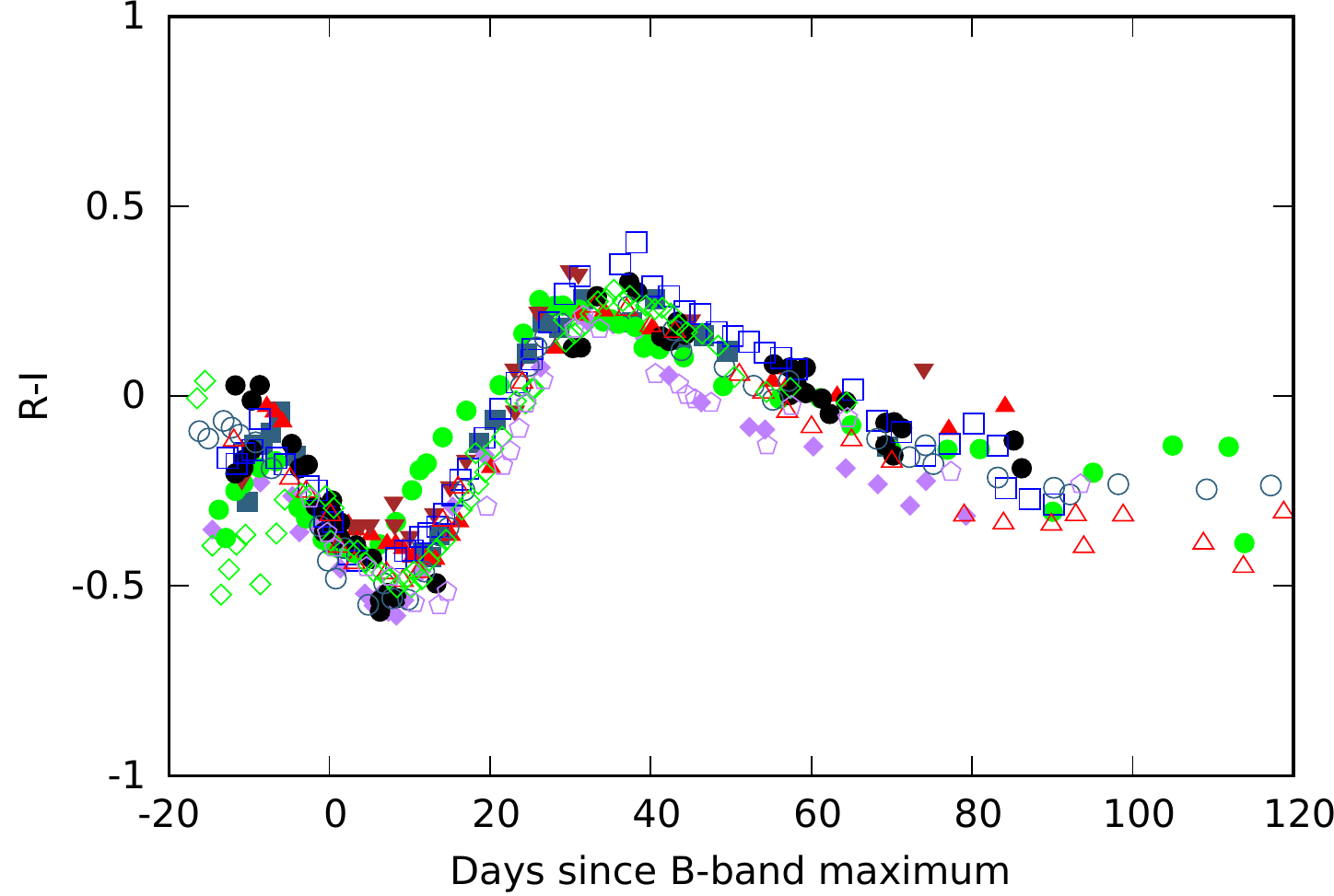}
   \end{subfigure}
   \caption{The $B-V$, $V-R$, $V-I$ and $R-I$ color curves of SN 2017fgc all together with comparison SNe Ia. All the peak colors are artificially shifted to match the observed values of SN 2017fgc at $t_{Bmax}$. Filled symbols represent HV SNe Ia, while open symbols represent NV SNe Ia.}
   \label{fig:CC}
 \end{figure*}

The $B-V$, $V-R$, $V-I$, and $R-I$ color curves of SN 2017fgc and the other SNe Ia are presented in Fig.~\ref{fig:CC}. All the color curves are shifted to match the peak color and the maximum epoch in the corresponding band of SN 2017fgc. The $B-V$ color evolution is similar for all the SNe Ia in the early epoch where the $B-V$ color rises for $\sim$ 30 days. As previously shown by \citet{2019ApJ...882..120W}, the $B-V$ color becomes bluer for HV SNe Ia for later epochs, where SN 2017fgc also shows the same behavior. The $B-V$ color evolution of SN 2017fgc is especially similar to SN 2002bo at all the epochs. 

The $V-R$ color curve does not show any notable change in the color at $\sim$ 20$-$40 days. $V-R$ color of SN 2017fgc follows a redder path, mainly following a close match to SN 2002bo. The $V-I$ color curve of SN 2017fgc shows a similar behavior until $\sim$ 30 days after the $B$-band maximum, after which SN 2017fgc follows a redder evolution.

Overall, the color evolution in $B-V$ and $V-R$ of SN 2017fgc shows a close match to those of SN 2002bo. In the $V-I$ and $R-I$ colors, SN 2017fgc follows a redder evolution after the peak. We do not see a clear difference between HV and NV SNe Ia around the secondary peaks. Given that the difference around the secondary peaks can be seen individually in the light curves in each band, this suggests that the excess might be caused by the differences in the bolometric luminosities (or at least the quasi-bolometric luminosities within the optical bands). We further discuss the property of the excess in the next section.

\section{Analysis of the Excess}
\label{sec:excess}
The light curve analysis of SN 2017fgc has revealed a characteristic shoulder in the $VRI$ bands, as shown in Figs.~\ref{fig:RBAND}$-$~\ref{fig:VBAND}. 

\subsection{Methodology}
In order to quantitatively study the difference between NV and HV SNe Ia light curves regarding the excess in the $VRI$ bands, we have used SN 2004dt \citep{2010ApJS..190..418G} and SN 2009ig \citep{2012ApJ...744...38F} and the same comparison objects previously used in this paper. We used the `max model' in SNooPy, which generates the templates from \citet{2006ApJ...647..501P} to fit individual light curves for all the comparison SNe. For each band, we first fit the light curve by omitting the data around the secondary peak, and obtain a light curve template in each band. This template predicts the light curve behavior around the secondary peak for the same object. Then, we subtract the entire light curve by the template for each object. The difference here is then defined as the amount of the excess. Omitting the data in the same epochs for HV and NV SNe Ia, and by setting the fit parameters free, we remove a possible bias from our calculations.

\begin{figure}
	 \centering
	 \includegraphics[width=\columnwidth]{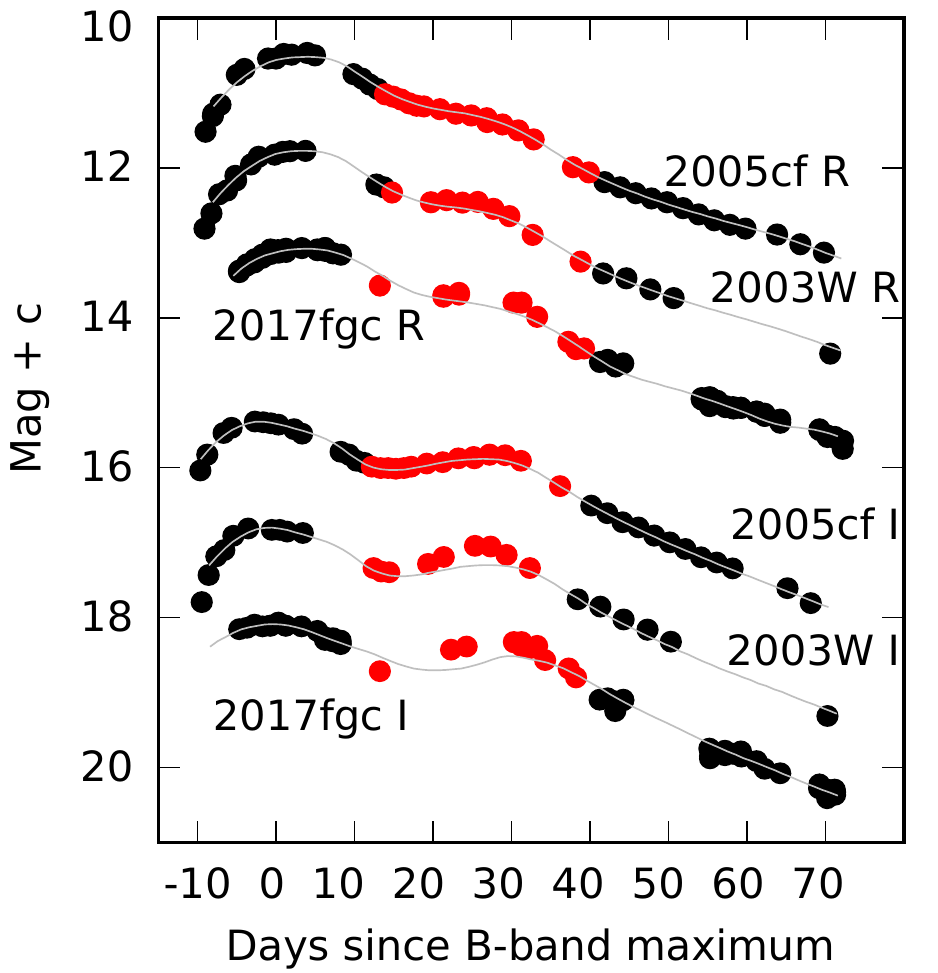}
     \caption{The comparison of the light curves of SN 2017fgc (HV), SN 2005cf (NV), and SN 2003W (HV). Gray line represents the fit obtained by using only the black dots. Red dots represent the data omitted in the fit. All data are shifted in the y-axis for clarity. The light curves are shifted to match the time of $B$-band maximum.}
     \label{fig:CompExcess}
	 \end{figure}

Fig.~\ref{fig:CompExcess} shows the examples of the procedure, for SN 2017fgc (HV), SN 2003W (HV), and SN 2005cf (NV). The template obtained by omitting the data around the secondary peak is shown by a gray line. It is seen that the template fits the secondary peak seen in NV SN 2005cf. On the other hand, HV SN 2017fgc and 2003W show the excess both in the $R$ and $I$ bands. The same procedure is applied to all other comparison SNe as well.

	 \begin{table*}
	 \setlength{\tabcolsep}{18pt}
	\caption{Maximum excess in each band for NV and HV SNe Ia.}
	\label{tab:excess}
	\centering
	\begin{tabular}{ccccc} 
		\hline
	Group & Name & V$_{excess}$ & R$_{excess}$ & I$_{excess}$\\
	 & & $(mag)$ & $(mag)$ & $(mag)$\\
		\hline
	NV & SN 2005cf & 0.02 $\pm$ 0.02 & 0.05 $\pm$ 0.03 & $-$0.08 $\pm$ 0.05\\
	NV & SN 2007af & $-$0.03 $\pm$ 0.02 & 0.12 $\pm$ 0.03 & 0.04 $\pm$ 0.02\\
	NV & SN 2011fe & 0.01 $\pm$ 0.02 & 0.09 $\pm$ 0.02 & $-$0.09 $\pm$ 0.04\\
	NV & SN 2005na & 0.08 $\pm$ 0.03 & 0.10 $\pm$ 0.03 & 0.06 $\pm$ 0.04\\
	NV & SN 2012cg & $-$0.05 $\pm$ 0.02 & $-$0.08 $\pm$ 0.03 & 0.06 $\pm$ 0.04\\
		\hline
	HV & SN 2017fgc & $-$0.21 $\pm$ 0.07 & $-$0.18 $\pm$ 0.04 & $-$0.28 $\pm$ 0.05\\
	HV & SN 2003W & $-$0.06 $\pm$ 0.02 & $-$0.15 $\pm$ 0.05 & $-$0.25 $\pm$ 0.06\\
	HV & SN 2007le & $-$0.15 $\pm$ 0.03 & $-$0.08 $\pm$ 0.02 & $-$0.15 $\pm$ 0.02\\
	HV & SN 2002bo & $-$0.06 $\pm$ 0.02 & $-$0.25 $\pm$ 0.05 & $-$0.15 $\pm$ 0.05\\
	HV & SN 2006X & $-$0.12 $\pm$ 0.03 & $-$0.17 $\pm$ 0.04 & $-$0.25 $\pm$ 0.04\\
	HV & SN 2002dj & $-$0.13 $\pm$ 0.04 & $-$0.06 $\pm$ 0.02 & $-$0.15 $\pm$ 0.03\\
	HV & SN 2004dt & $-$0.17 $\pm$ 0.03 & $-$0.24 $\pm$ 0.03 & $-$0.45 $\pm$ 0.05\\
	HV & SN 2009ig & $-$0.14 $\pm$ 0.06 & $-$0.14 $\pm$ 0.05 & $-$0.18 $\pm$ 0.06\\
		\hline
	\end{tabular}
\end{table*}

	 \begin{figure}
	 \centering
	 \includegraphics[width=\columnwidth]{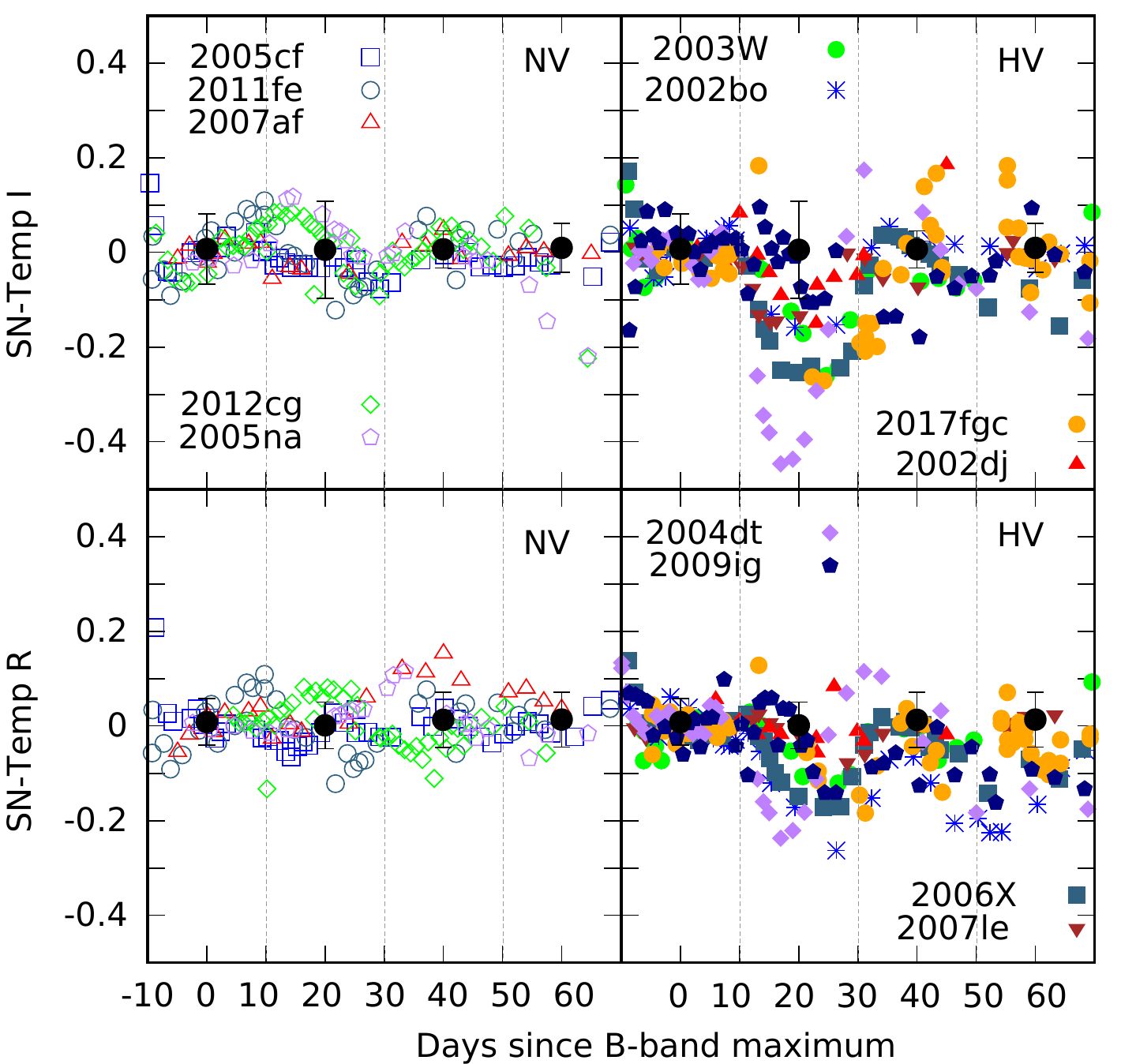}
     \caption{The difference between the template and observed light curves. Filled symbols represent HV SNe Ia, while open symbols represent NV SNe Ia. The top panels are for the $I$-band data, while the bottom panels are for the $R$-band data. The black dots are the average values for the NV SNe Ia samples within four different time bins, with the associated statistical scatters. In order to clarify the difference in magnitudes, 
    the same average values for the `NV SNe Ia sample' are shown in the panels for the HV SNe Ia. SN 2017fgc is represented with orange color.}
     \label{fig:TempDaysMax}
	 \end{figure} 

\subsection{Property of the excess}
The results obtained with the same procedure for the entire sample is shown in Fig.~\ref{fig:TempDaysMax}. It quantitatively shows that at $\sim$ 20 days there is a distinct difference between HV and NV SNe Ia. NV SNe Ia show a uniform nature. HV SNe Ia show a large scatter, clearly exceeding the statistical fluctuation seen in NV SNe Ia. Average values of the excess ($\sim$0 mag) obtained for 20 days intervals from NV SNe Ia data are also shown for both NV and HV SNe Ia samples to make the comparison easier.

While there is a small diversity in the behavior of HV SNe Ia in the epoch and the magnitude of the excess, the same tendency is seen. They all show an excess compared to NV SNe Ia at similar epochs. This excess tends to be stronger in redder bands. Table~\ref{tab:excess} shows the maximum excess seen in the $VRI$ bands for the comparison SNe.

\begin{figure}
   \centering
   \begin{subfigure}[b]{0.49\linewidth}
     \includegraphics[width=\linewidth]{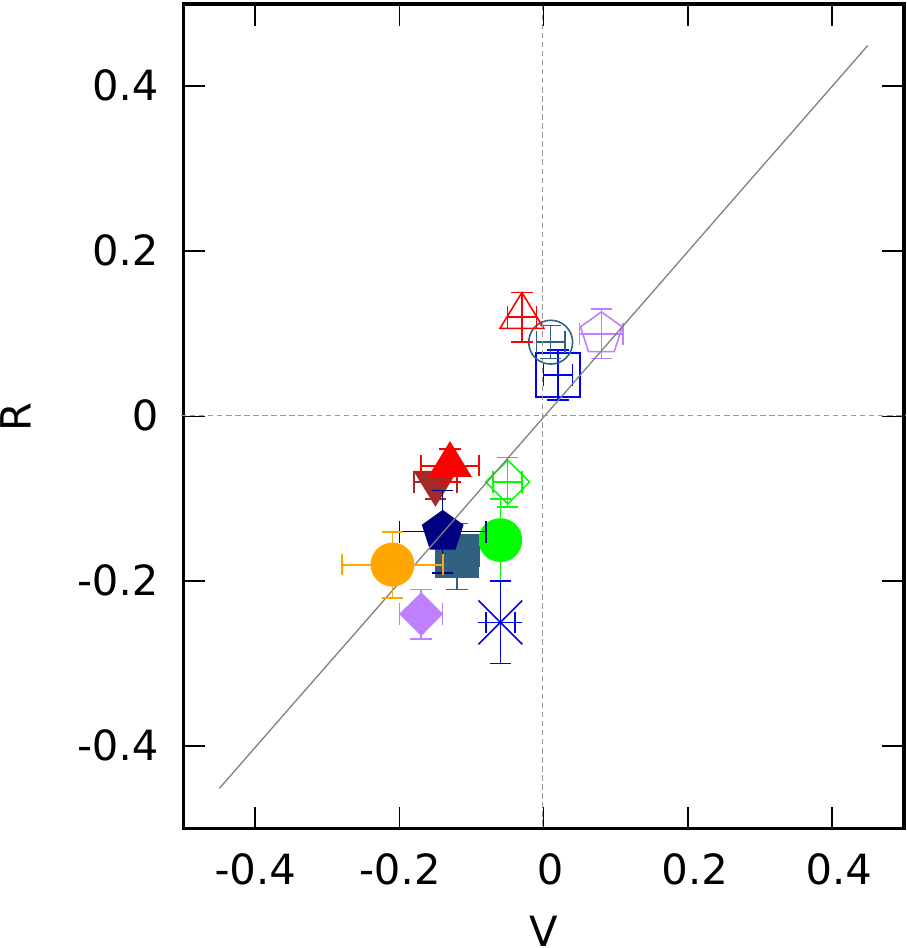}
   \end{subfigure}
   \begin{subfigure}[b]{0.49\linewidth}
     \includegraphics[width=\linewidth]{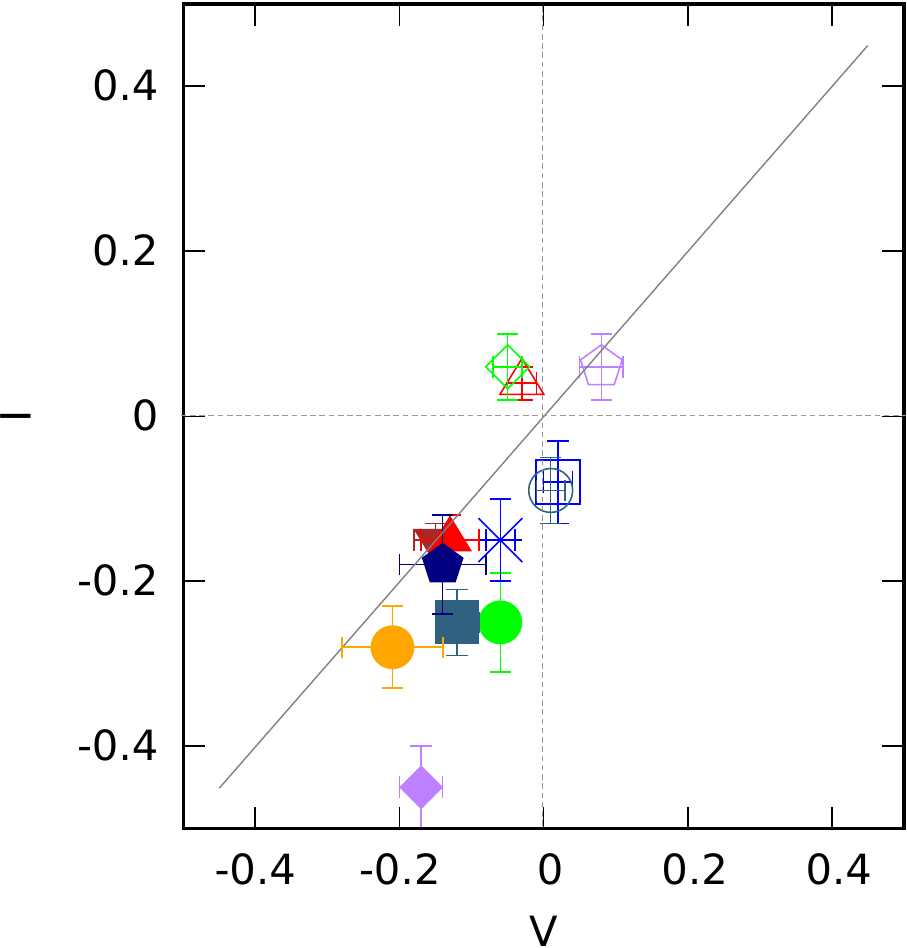}
   \end{subfigure}
      \begin{subfigure}[b]{0.49\linewidth}
     \includegraphics[width=\linewidth]{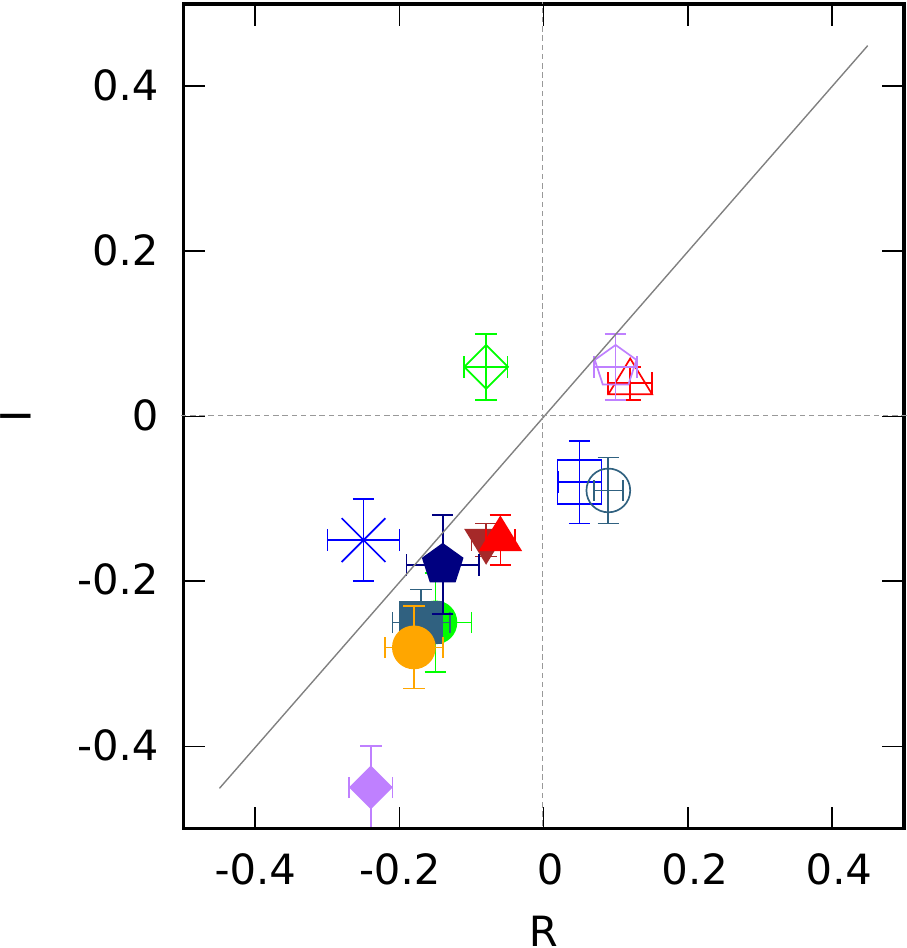}
   \end{subfigure}
   \caption{The excess in each band. The same symbols as in  Fig.~\ref{fig:TempDaysMax} are used. Filled symbols represent HV SNe Ia and open symbols represent NV SNe Ia. Gray solid lines represent the lines for $R-V$=0, $I-V$=0, and $I-R$=0. SN 2017fgc is represented with orange color, while the blue star sign represents SN 2002bo.}
   \label{fig:SneTempColor}
 \end{figure}

The excess in each band is compared in Fig.~\ref{fig:SneTempColor}. NV SNe Ia have no clear excess, being clustered around the zero points. HV SNe Ia have a large scatter showing an excess, which is stronger in the redder bands in general. 

\section{Discussion and Conclusions}

In this paper, we present the extensive photometric and spectroscopic observations of SN 2017fgc, which spans from 12 days before the $B$-band maximum to 137 days afterward. SN 2017fgc is characterized by a normal decline rate of $\Delta m _{15} (B)_{true}$=1.10 $\pm$ 0.10 mag. We found the reddening free peak absolute magnitude of SN 2017fgc as $M_B$ = $-$19.34$ \pm $ 0.19 mag. We present the relevant parameters obtained for SN 2017fgc in Table~\ref{tab:parameters}. Spectroscopically, SN 2017fgc is a HV SN Ia with an expansion velocity of $\sim$ 20,000 km $s^{-1}$ at 13 days prior to the $B$-band maximum, and $\sim 15,200$ km s$^{-1}$ around the maximum. Spectra of SN 2017fgc well match with those of SN 2002bo in all the epochs. 

We have compared the multiband light curves of SN 2017fgc to those of NV SNe Ia, which led us to find a distinct behavior. SN 2017fgc shows a shoulder-like behaviour in the $V$ band, and an excess ($\sim$ 0.2 mag level) in the $R$ and $I$ bands around the NIR secondary peaks are found. This is similar to the behaviour previously noticed for SN 2006X \citep{2008ApJ...675..626W} and SN 2002dj \citep{2008MNRAS.388..971P}. We have further found that this is a common property for HV SNe Ia in general in the $VRI$ bands.

We have studied the property of the excess in the $R$ and $I$ bands. By fitting the light curves of individual SNe by omitting the data near the NIR secondary peak we have obtained the light curve templates in each band. Then those templates were used to predict the secondary peaks. Comparing the templates and the observed light curves of HV SNe Ia and NV SNe Ia, we have found that the templates are in good agreement with NV SNe Ia. On the other hand, we found a difference between the light curves of HV SNe Ia and the templates due to the pronounced shoulder, which revealed an excess in HV SNe Ia that is stronger in the redder bands. 

As a possible origin of the excess, one might consider an external effect caused by the environment. \citet{2009ApJ...693..207B} studied the NaI D variability in highly reddened SNe Ia which tend to be HV SNe Ia. The variability in NaI D raises the possibility that the SN light could affect the environment in a time-dependent way by destructing the surrounding dust, which will reduce extinction for later epochs. However, in the V, R and I band light curves, HV and NV SNe Ia go back to the same track after $\sim$40 days. Therefore, a possible change in the extinction property would not be an origin of the excess found in our study.

If a dusty environment (e.g., within a circumstellar environment) exists near the SN, an echo might be created that will contribute to the light curves as an additional component. In the NIR, this is caused by a thermal emission echo \citep{2015MNRAS.452.3281M}. For example, \citet{2017ApJ...835..143N} investigated the distribution of the circumstellar dust around SN 2012dn and explained the NIR excess starting around 30 days after the $B$-band maximum by this thermal emission echo. This scenario naturally predicts a blue echo in the optical as created by the dust scatterings. Indeed, \citet{2019ApJ...882..120W} suggested the scattering light echo as an origin of the late time excess after $\sim 30$ days since the $B$-band maximum (see Section 4), which is seen only in the short wavelengths and matches to the echo prediction. However, the behavior found in our study is different; the excess here takes place much earlier, and it is red. We, therefore, conclude that the excess is not caused by the CS echo. In summary, it is very unlikely that the excess is attributed to any external effect.

The origin of the NIR secondary peak has not been fully understood, but it is likely that this is caused by the recombination of Fe III to Fe II in the ejecta \citep{2006ApJ...649..939K}. Since this is caused by the existence of iron in the ejecta, our finding would indicate that HV SNe Ia tend to have a more extended distribution of $^{56}$Ni and Fe-peak elements in the ejecta than in NV SNe Ia. One may expect that the difference in the ionization status would also be caused by the difference in the luminosity between HV and NV SNe Ia even without an intrinsic difference in the ejecta structure, since the HV and NV SNe Ia may have a small offset in the luminosity for a given decline rate. However, this effect is probably small; we find the excess exist in the sample of HV SNe Ia with a range of the decline rate, and therefore any small difference in the luminosity between HV and NV SNe Ia (with similar decline rates) should not be the main cause.

\citet{2020ApJ...895L...5P} suggested that the metallicity is likely to be the dominant factor in the creation of HV SNe Ia, which was initially proposed by \citet{2013Sci...340..170W}. Indeed, \citet{2000ApJ...530..966L} suggested that metallicity may affect the Si II $\lambda$ 6355 velocity, strength, and profile. However, we note that this direct effect of high abundance of iron due to high metallicty on the line formation will not provide good explanation for the prominent secondary feature observed in HV SNe Ia. Since the primordial iron within the ejecta will be distributed uniformly, it is expected for the ionization to work in a similar way as mixing of iron within the ejecta. Indeed, a larger degree of mixing would suppress the NIR secondary peak \citep{2006ApJ...649..939K}, which is opposite to the observed behavior.

Irrespective of a cause in the different ionization levels, our finding supports the existence of a difference seen in the ejecta between HV and NV SNe Ia. Indeed, \citet{2020ApJ...893..143K} performed a spectral synthesis study for HV SN 2019ein and argued that there is an intrinsic difference in the ejecta properties between HV and NV SNe Ia, where HV SNe Ia have a more extended distribution of burning products. 

Given that NV and HV SNe Ia share similar luminosity range (i.e., $^{56}$Ni), it is difficult to explain the difference in the distribution of $^{56}$Ni by a single model sequence if the model would be controlled by a single parameter. The popular model, a delayed-detonation on a nearly Chandrasekhar-mass WD, is mainly characterized by the central density and the deflagration-to-detonation transition density; these may be set through the accretion history \citep{1984ApJ...286..644N} and the C/O ratio in the progenitor \citep{1999ApJ...522L..43U}, respectively. Yet another popular model, a double-detonation on a sub-Chandrasekhar-mass WD, can be characterized by the masses of the WD and the He shell; different combinations on these parameters are possible depending on the evolutionary pathway \citep{2011ApJ...734...38W}. Both scenarios can lead to an asymmetric distribution of burning products including $^{56}$Ni \citetext{\citealp{2010ApJ...712..624M}; \citealp{2010A&A...514A..53F}}, which could contribute to explain at least part of the HV/NV SNe Ia diversity by a viewing angle effect \citep{2010Natur.466...82M}. Given systematically different environments found for NV and HV SNe Ia \citep{2013Sci...340..170W}, it is likely that the SNe Ia progenitors can be a mixture of different progenitor systems leading to different explosion models. While further discussion on the origin of the difference in the ejecta property is beyond the scope of the presented work, the results presented in this work highlight the importance of high-quality and well-sampled light curve data for a large sample of HV SNe Ia.

\label{sec:conc}

\begin{table}
\centering
\begin{minipage}[b]{\columnwidth}
	\caption{Relevant parameters for SN 2017fgc and its host galaxy}
	\label{tab:parameters}
	\begin{tabular*}{\textwidth}{l@{\extracolsep{\fill}}cc} 
		\hline
		Parameter & Value & Source \\
		\hline
	     & SN 2017fgc & \\
				\\
		Discovery date UT & 2017 July 11.16 & 1\\
		Discovery MJD & 57945.47 & 1\\
		Discovery mag & 17.32 mag & 1\\
		Epoch of B maximum & 57958.721 $\pm$ 0.693 & 2\\
		$B_{max}$ & 13.97$\pm$0.02 mag & 2\\
		$B_{max}$ $-$ $V_{max}$ & 0.21$\pm$0.03 mag & 2\\
		$M_{Bmax}$ & $-$19.34 $ \pm $ 0.19 mag & 2\\
		E(B$-$V)$_{host}$ & 0.29 $\pm$ 0.02 mag & 2\\
		$ \Delta m_{15}(B)_{observed}$& 1.08 $ \pm $ 0.09 mag& 2\\
		$ \Delta m_{15}(B)_{true}$& 1.10 $ \pm $ 0.10 mag & 2\\
		\\
		  & NGC 0474 & \\
				\\
		Galaxy type & $SA0^0(s)$ & 3\\
		E(B$-$V)$_{Gal}$ & 0.029 & 3\\
		$B_{tot}$ & 12.38 & 4\\
		$v_{vir}$ & 2321 km s$^{-1}$ & 3\\
		$v_{3k}$ & 2001 km s$^{-1}$ & 3\\
		Distance & 27.7 Mpc & 4\\
		$\mu$ & 32.35$\pm$0.15 & 3\\
		\hline
	\end{tabular*}
	       {\raggedright} References: 1 = Valenti et al. (2017); 2 = this paper; 3 = NASA Extragalactic Database (http://ned.ipac.caltech.edu); 4 = Kim et al. (2012)
	\end{minipage}
\end{table} 

\section*{Acknowledgements}

We acknowledge ESA Gaia, DPAC and the Photometric Science Alerts Team (http://gsaweb.ast.cam.ac.uk/alerts). U.B. acknowledges the support provided by the Turkish Scientific and Technical Research Council (T\"UB\.ITAK$-$2211C and 2214A). The authors thank T\"UB\.ITAK for a partial support in using T60 telescope with the project number 17BT60-1185. K.M. acknowledges support by JSPS KAKENHI Grant (JP20H00174, JP20H0473, JP18H04585, JP18H05223, JP17H02864). B.K. acknowledges support by T\"UB\.ITAK (Project no:120F169). The authors thank the anonymous referee for her/his valuable comments and suggestions that have significantly improved the manuscript.

\section*{DATA AVAILABILITY}

The data underlying this article will be shared on reasonable request to the corresponding author.






\appendix

\section{Photometry Table and Spectra Comparisons}

\begin{table*}
	\centering
	\caption{Photometric observations of SN 2017fgc.}
	\label{tab:tugdata}
	\begin{tabular*}{\textwidth}{c @{\extracolsep{\fill}} cccccc}
		\hline
		Date & $MJD^a$ & $Phase^b$ & B & V & R & I\\
		 & & $(d)$ & $(mag)$ & $(mag)$ & $(mag)$ & $(mag)$\\
		\hline
2017 Jul 13&57947.0&$-$11.7&15.92 $\pm$ 0.05&15.61 $\pm$ 0.03 &15.34 $\pm$ 0.04&15.32 $\pm$ 0.06\\
 &57947.0&$-$11.7&15.97 $\pm$ 0.05&...&15.31 $\pm$ 0.04&15.52 $\pm$ 0.08\\
 &57947.0&$-$11.7&...&...&15.34 $\pm$ 0.03 &15.32 $\pm$ 0.07\\
 &57947.0&$-$11.7&...&...&...&15.52 $\pm$ 0.08\\
2017 Jul 14&57948.0&$-$10.7&15.13 $\pm$ 0.04&15.21 $\pm$ 0.03 &14.97 $\pm$ 0.03 &15.14 $\pm$ 0.06\\
 &57948.0&$-$10.7&...&15.26 $\pm$ 0.03 &14.93 $\pm$ 0.03 &15.12 $\pm$ 0.07\\
 &57948.0&$-$10.7&...&...&14.96 $\pm$ 0.03 &...\\
2017 Jul 15&57949.0&$-$9.7&15.01 $\pm$ 0.04&14.99 $\pm$ 0.03 &14.75 $\pm$ 0.03 &14.90 $\pm$ 0.07\\
 &57949.0&$-$9.7&...&...&14.83 $\pm$ 0.03 &14.84 $\pm$ 0.08\\
2017 Jul 16&57950.0&$-$8.7&...&14.80 $\pm$ 0.05&...&14.62 $\pm$ 0.05\\
 &57950.0&$-$8.7&...&14.84 $\pm$ 0.05&14.65 $\pm$ 0.04&...\\
2017 Jul 20&57954.0&$-$4.7&...&...&13.88 $\pm$ 0.02 &...\\
 &57954.0&$-$4.7&14.23 $\pm$ 0.03 &14.05 $\pm$ 0.02 &13.87 $\pm$ 0.02 &14.00 $\pm$ 0.05\\
 &57954.0&$-$4.7&...&14.09 $\pm$ 0.02 &13.89 $\pm$ 0.03 &...\\
 &57954.0&$-$4.7&...&14.12 $\pm$ 0.03 &...&...\\
 &57954.0&$-$4.7&...&14.09 $\pm$ 0.02 &...&...\\
2017 Jul 21&57955.0&$-$3.7&14.07 $\pm$ 0.03 &13.95 $\pm$ 0.02 &13.79 $\pm$ 0.03 &13.98 $\pm$ 0.05\\
 &57955.0&$-$3.7&...&...&13.80 $\pm$ 0.02 &13.94 $\pm$ 0.05\\
2017 Jul 22&57956.0&$-$2.7&...&...&13.76 $\pm$ 0.02 &...\\
 &57956.0&$-$2.7&...&13.90 $\pm$ 0.02 &13.72 $\pm$ 0.02 &...\\
 &57956.0&$-$2.7&...&13.94 $\pm$ 0.02 &13.74 $\pm$ 0.02 &...\\
2017 Jul 23&57957.0&$-$1.7&...&13.82 $\pm$ 0.02 &13.70 $\pm$ 0.02 &...\\
 &57957.0&$-$1.7&14.08 $\pm$ 0.03 &13.85 $\pm$ 0.02 &13.67 $\pm$ 0.02 &13.96 $\pm$ 0.05\\
2017 Jul 24&57958.0&$-$0.7&...&13.79 $\pm$ 0.02 &13.60 $\pm$ 0.02 &13.95 $\pm$ 0.05\\
 &57958.0&$-$0.7&...&...&13.65 $\pm$ 0.02 &...\\
 &57958.0&$-$0.7&...&13.80 $\pm$ 0.02 &13.63 $\pm$ 0.02 &13.95 $\pm$ 0.05\\
2017 Jul 26&57960.0&1.3&14.03 $\pm$ 0.03 &...&13.61 $\pm$ 0.02 &13.92 $\pm$ 0.05\\
 &57960.0&1.3&...&13.73 $\pm$ 0.02 &13.63 $\pm$ 0.02 &13.91 $\pm$ 0.05\\
 &57960.0&1.3&...&...&13.58 $\pm$ 0.02 &13.95 $\pm$ 0.05\\
 &57960.0&1.3&...&...&...&13.95 $\pm$ 0.05\\
 &57960.0&1.3&...&13.75 $\pm$ 0.02 &13.62 $\pm$ 0.02 &...\\
 &57960.0&1.3&...&13.73 $\pm$ 0.02 &...&...\\
2017 Jul 28&57962.0&3.3&14.12 $\pm$ 0.03 &13.74 $\pm$ 0.02 &13.57 $\pm$ 0.02 &13.96 $\pm$ 0.05\\
 &57962.0&3.3&...&13.64 $\pm$ 0.02 &...&...\\
 &57962.0&3.3&...&13.64 $\pm$ 0.02 &...&...\\
2017 Jul 30&57964.0&5.2&14.25 $\pm$ 0.03 &13.67 $\pm$ 0.02 &13.59 $\pm$ 0.02 &14.02 $\pm$ 0.05\\
 &57964.0&5.3&...&...&13.60 $\pm$ 0.02 &...\\
2017 Jul 31&57965.0&6.2&14.37 $\pm$ 0.03 &13.67 $\pm$ 0.02 &13.58 $\pm$ 0.02 &14.11 $\pm$ 0.05\\
 &57965.0&6.2&...&13.69 $\pm$ 0.02 &13.58 $\pm$ 0.02 &14.14 $\pm$ 0.06\\
 &57965.0&6.2&...&...&13.61 $\pm$ 0.02 &...\\
2017 Aug 01&57966.0&7.2&14.41 $\pm$ 0.03 &...&13.64 $\pm$ 0.02 &14.16 $\pm$ 0.05\\
 &57966.0&7.3&14.33 $\pm$ 0.03 &...&...&14.12 $\pm$ 0.05\\
2017 Aug 02&57967.0&8.2&14.42 $\pm$ 0.03 &13.78 $\pm$ 0.02 &13.66 $\pm$ 0.02 &14.19 $\pm$ 0.06\\
 &57967.0&8.2&...&...&...&14.15 $\pm$ 0.05\\
 &57967.0&8.2&...&...&...&14.20 $\pm$ 0.05\\
2017 Aug 07&57972.0&13.2&15.04 $\pm$ 0.05&...&14.07 $\pm$ 0.03 &14.56 $\pm$ 0.05\\
 &57972.0&13.2&15.06 $\pm$ 0.05&...&...&...\\
2017 Aug 15&57980.1&21.3&...&14.34 $\pm$ 0.02 &14.20 $\pm$ 0.02 &...\\
 &57980.1&21.3&...&...&14.22 $\pm$ 0.02 &...\\
2017 Aug 16&57981.0&22.3&...&...&...&14.27 $\pm$ 0.05\\
2017 Aug 17&57982.1&23.3&...&14.39 $\pm$ 0.02 &14.19 $\pm$ 0.02 &...\\
 &57982.1&23.3&...&...&14.17 $\pm$ 0.02 &...\\
2017 Aug 18&57983.0&24.3&...&...&...&14.23 $\pm$ 0.05\\
2017 Aug 22&57987.0&28.3&...&14.51 $\pm$ 0.06&...&...\\
2017 Aug 24&57989.0&30.3&...&...&14.30 $\pm$ 0.02 &...\\
 &57989.0&30.3&16.68 $\pm$ 0.04&14.73 $\pm$ 0.02 &14.30 $\pm$ 0.02 &14.17 $\pm$ 0.04\\
2017 Aug 25&57990.0&31.3&...&...&...&14.17 $\pm$ 0.05\\
 &57990.0&31.3&...&14.78 $\pm$ 0.02 &...&14.20 $\pm$ 0.05\\
 &57990.1&31.3&...&...&...&14.23 $\pm$ 0.05\\
2017 Aug 26&57991.1&32.3&...&14.90 $\pm$ 0.03 &14.49 $\pm$ 0.02 &14.25 $\pm$ 0.05\\
2017 Aug 27&57992.0&33.3&...&...&...&14.22 $\pm$ 0.05\\
2017 Aug 28&57993.0&34.3&...&...&...&14.41 $\pm$ 0.05\\
		\hline
	\end{tabular*}
      \\ {\raggedright   ($^a$)MJD=JD-2400000.5, ($^b$)Relative to $B$-band maximum (MJD = 57958.721) \par}
\end{table*}

\begin{table*}
	\centering
	\contcaption{Photometric observations of SN 2017fgc.}
	\label{tab:continued}
	\begin{tabular*}{\textwidth}{c @{\extracolsep{\fill}} cccccc}
		\hline
		Date & $MJD^a$ & $Phase^b$ & B & V & R & I\\
		 & & $(d)$ & $(mag)$ & $(mag)$ & $(mag)$ & $(mag)$\\
		\hline
2017 Aug 31&57996.0&37.3&...&15.29 $\pm$ 0.02 &14.82 $\pm$ 0.02 &...\\
 &57996.0&37.3&...&...&14.92 $\pm$ 0.02 &14.52 $\pm$ 0.05\\
2017 Sept 01&57997.0&38.3&...&...&...&14.64 $\pm$ 0.05\\
2017 Sept 02&57998.0&39.3&...&...&14.91 $\pm$ 0.02 &...\\
2017 Sept 04&58000.0&41.3&...&15.51 $\pm$ 0.04&15.09 $\pm$ 0.02 &14.94 $\pm$ 0.07\\
2017 Sept 05&58001.0&42.2&...&...&15.07 $\pm$ 0.02 &14.92 $\pm$ 0.06\\
2017 Sept 06&58002.0&43.2&...&15.68 $\pm$ 0.05&15.15 $\pm$ 0.03 &14.98 $\pm$ 0.05\\
2017 Sept 07&58003.0&44.2&...&15.53 $\pm$ 0.04&15.11 $\pm$ 0.03 &14.94 $\pm$ 0.04\\
2017 Sept 17&58013.0&54.2&17.10 $\pm$ 0.04&15.90 $\pm$ 0.02 &15.58 $\pm$ 0.03 &...\\
 &58013.0&54.3&...&...&15.59 $\pm$ 0.03 &...\\
 &58013.0&54.3&17.03 $\pm$ 0.04&...&15.68 $\pm$ 0.04&15.59 $\pm$ 0.05\\
 &58013.0&54.3&17.13 $\pm$ 0.04&...&15.56 $\pm$ 0.03 &...\\
2017 Sept 18&58014.0&55.3&...&...&15.56 $\pm$ 0.02 &...\\
 &58014.0&55.3&...&15.95 $\pm$ 0.03 &15.58 $\pm$ 0.03 &15.69 $\pm$ 0.06\\
2017 Sept 19&58015.0&56.3&...&16.00 $\pm$ 0.03 &15.61 $\pm$ 0.03 &15.69 $\pm$ 0.05\\
 &58015.0&56.3&17.12 $\pm$ 0.04&15.95 $\pm$ 0.03 &15.68 $\pm$ 0.02 &...\\
 &58015.0&56.3&17.17 $\pm$ 0.04&15.91 $\pm$ 0.03 &15.69 $\pm$ 0.03 &...\\
 &58015.0&56.3&17.18 $\pm$ 0.04&15.97 $\pm$ 0.03 &...&...\\
2017 Sept 20&58016.0&57.3&...&15.96 $\pm$ 0.03 &...&...\\
 &58016.0&57.3&...&15.98 $\pm$ 0.03 &...&...\\
 &58016.0&57.3&17.16 $\pm$ 0.04&15.99 $\pm$ 0.03 &15.69 $\pm$ 0.03 &15.66 $\pm$ 0.06\\
 &58016.0&57.3&17.24 $\pm$ 0.04&15.99 $\pm$ 0.03 &15.71 $\pm$ 0.03 &...\\
 &58016.0&57.3&17.11 $\pm$ 0.04&16.01 $\pm$ 0.03 &...&...\\
2017 Sept 21&58017.0&58.3&17.15 $\pm$ 0.04&16.02 $\pm$ 0.03 &15.70 $\pm$ 0.03 &15.63 $\pm$ 0.05\\
 &58017.0&58.3&17.28 $\pm$ 0.04&16.02 $\pm$ 0.03 &15.70 $\pm$ 0.03 &...\\
 &58017.9&59.2&17.20 $\pm$ 0.04&16.03 $\pm$ 0.04&...&...\\
2017 Sept 22&58018.0&59.3&...&16.03 $\pm$ 0.03 &...&15.70 $\pm$ 0.05\\
2017 Sept 23&58019.9&61.2&17.26 $\pm$ 0.04&16.07 $\pm$ 0.04&15.75 $\pm$ 0.03 &15.76 $\pm$ 0.06\\
 &58019.9&61.2&17.13 $\pm$ 0.03 &...&15.76 $\pm$ 0.02 &...\\
2017 Sept 24&58020.0&61.2&17.16 $\pm$ 0.04&...&...&...\\
 &58020.9&62.2&17.22 $\pm$ 0.03 &16.20 $\pm$ 0.03 &15.81 $\pm$ 0.02 &15.86 $\pm$ 0.07\\
2017 Sept 25 &58021.0&62.2&17.17 $\pm$ 0.03 &16.08 $\pm$ 0.04&15.78 $\pm$ 0.02 &...\\
 &58021.0&62.3&...&16.15 $\pm$ 0.05&15.81 $\pm$ 0.03 &...\\
 &58021.0&62.3&...&16.09 $\pm$ 0.04&...&...\\
2017 Sept 27&58023.0&64.2&17.19 $\pm$ 0.04&16.16 $\pm$ 0.04&15.91 $\pm$ 0.03 &15.92 $\pm$ 0.07\\
 &58023.0&64.2&17.20 $\pm$ 0.04&16.13 $\pm$ 0.05&15.86 $\pm$ 0.03 &...\\
 &58023.0&64.3&...&16.18 $\pm$ 0.06&...&...\\
2017 Oct 01&58027.9&69.2&17.21 $\pm$ 0.05&16.33 $\pm$ 0.03 &15.99 $\pm$ 0.03 &16.09 $\pm$ 0.05\\
 &58027.9&69.2&17.28 $\pm$ 0.04&16.25 $\pm$ 0.05&16.00 $\pm$ 0.03 &...\\
2017 Oct 02&58028.9&70.2&17.31 $\pm$ 0.06&16.39 $\pm$ 0.07&16.09 $\pm$ 0.03 &16.25 $\pm$ 0.09\\
2017 Oct 03&58029.0&70.3&...&16.30 $\pm$ 0.04&16.07 $\pm$ 0.03 &16.13 $\pm$ 0.06\\
 &58029.0&70.3&...&...&16.10 $\pm$ 0.03 &...\\
2017 Oct 04&58030.0&71.3&...&16.38 $\pm$ 0.04&16.09 $\pm$ 0.03 &16.18 $\pm$ 0.04\\
 &58030.0&71.3&...&...&...&...\\
 &58030.0&71.3&...&...&16.25 $\pm$ 0.03 &...\\
2017 Oct 05&58031.0&72.3&...&16.37 $\pm$ 0.04&16.14 $\pm$ 0.04&...\\
 &58031.0&72.3&...&16.41 $\pm$ 0.04&...&...\\
2017 Oct 17&58043.9&85.2&17.57 $\pm$ 0.05&16.71 $\pm$ 0.04&16.53 $\pm$ 0.03 &16.66 $\pm$ 0.05\\
2017 Oct 18&58044.9&86.2&...&...&16.66 $\pm$ 0.04&16.68 $\pm$ 0.05\\
2017 Oct 23&58049.9&91.1&17.66 $\pm$ 0.07&16.87 $\pm$ 0.05&...&...\\
2017 Oct 31&58057.9&99.1&17.73 $\pm$ 0.08&...&17.12 $\pm$ 0.03 &...\\
 &58057.9&99.2&...&...&17.12 $\pm$ 0.04&...\\
2017 Nov 07&58064.8&106.1&...&...&17.48 $\pm$ 0.04&...\\
2017 Nov 08&58065.8&107.1&17.77 $\pm$ 0.04&...&17.43 $\pm$ 0.03 &...\\
2017 Nov 09&58066.8&108.1&17.88 $\pm$ 0.06&17.43 $\pm$ 0.05&17.49 $\pm$ 0.07&...\\
2017 Nov 10&58067.8&109.1&...&...&17.46 $\pm$ 0.05&17.42 $\pm$ 0.04\\
2017 Nov 11&58068.8&110.1&...&17.42 $\pm$ 0.04&17.56 $\pm$ 0.05&...\\
2017 Nov 14&58071.8&113.1&...&...&17.56 $\pm$ 0.04&...\\
2017 Nov 15&58072.8&114.1&...&...&17.57 $\pm$ 0.05&...\\
 &58072.8&114.1&...&...&17.54 $\pm$ 0.04&...\\
2017 Nov 16&58073.8&115.1&18.29 $\pm$ 0.07&...&17.66 $\pm$ 0.05&17.51 $\pm$ 0.04\\
2017 Nov 18&58075.8&117.1&18.21 $\pm$ 0.06&...&...&...\\
		\hline
	\end{tabular*}
      \\ {\raggedright   ($^a$)MJD=JD-2400000.5, ($^b$)Relative to $B$-band maximum (MJD = 57958.721) \par}
\end{table*}

\begin{table*}
	\centering
	\contcaption{Photometric observations of SN 2017fgc.}
	\begin{tabular*}{\textwidth}{c @{\extracolsep{\fill}} cccccc}
		\hline
		Date & $MJD^a$ & $Phase^b$ & B & V & R & I\\
		 & & $(d)$ & $(mag)$ & $(mag)$ & $(mag)$ & $(mag)$\\
		\hline
2017 Nov 22&58079.8&121.1&...&...&17.91 $\pm$ 0.04&...\\
2017 Nov 24&58081.8&123.1&18.35 $\pm$ 0.06&17.71 $\pm$ 0.04&17.92 $\pm$ 0.05&...\\
 &58081.8&123.1&...&17.60 $\pm$ 0.04&...&...\\
2017 Dec 02&58089.8&131.1&...&17.78 $\pm$ 0.09&...&...\\
2017 Dec 08&58095.8&137.0&...&...&18.36 $\pm$ 0.05&...\\
		\hline
	\end{tabular*}
      \\ {\raggedright   ($^a$)MJD=JD-2400000.5, ($^b$)Relative to $B$-band maximum (MJD = 57958.721) \par}
\end{table*}

   \begin{figure*}
   \centering
   \begin{subfigure}[b]{0.4\linewidth}
     \includegraphics[width=\linewidth]{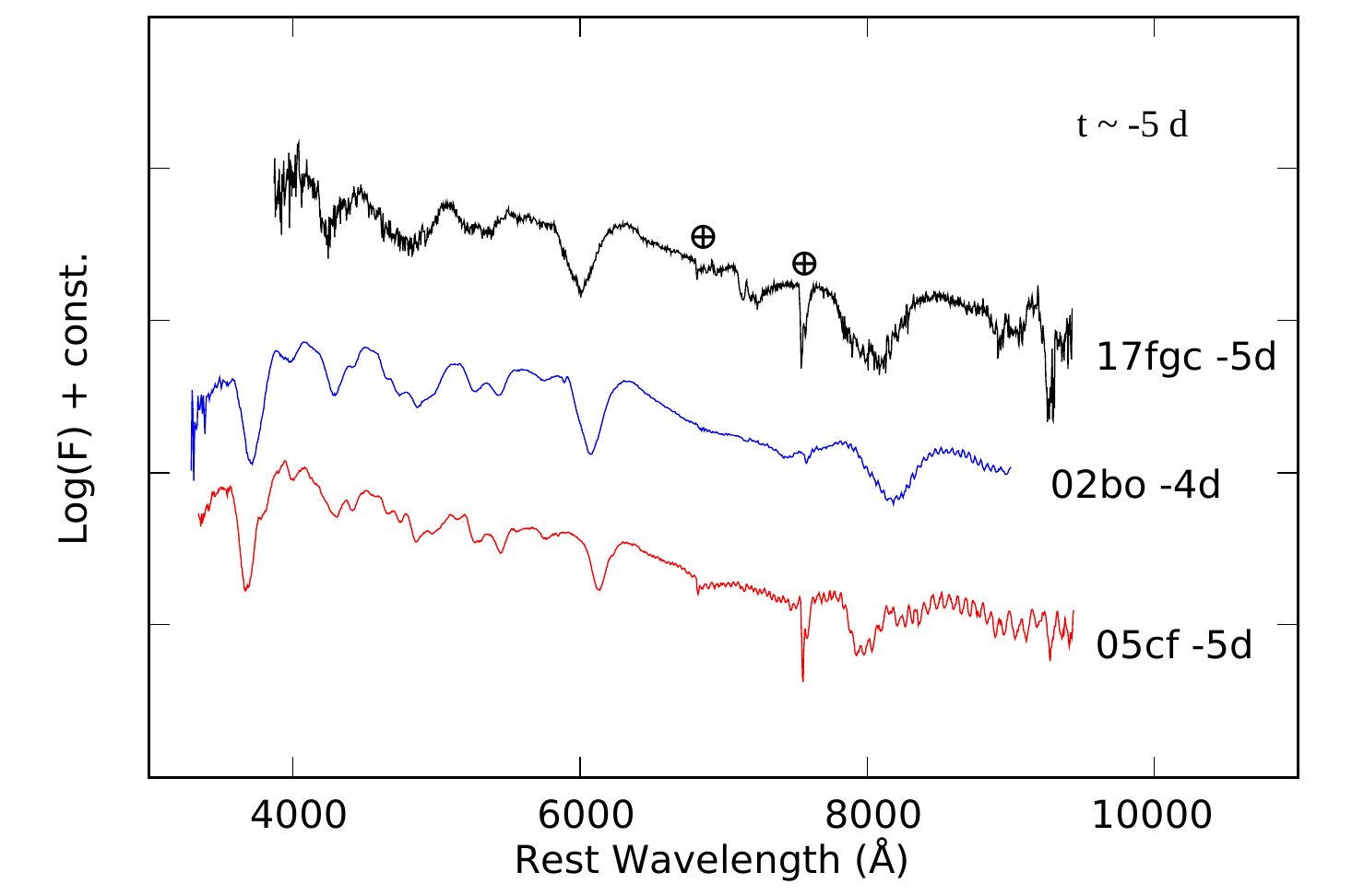}
   \end{subfigure}
   \begin{subfigure}[b]{0.4\linewidth}
     \includegraphics[width=\linewidth]{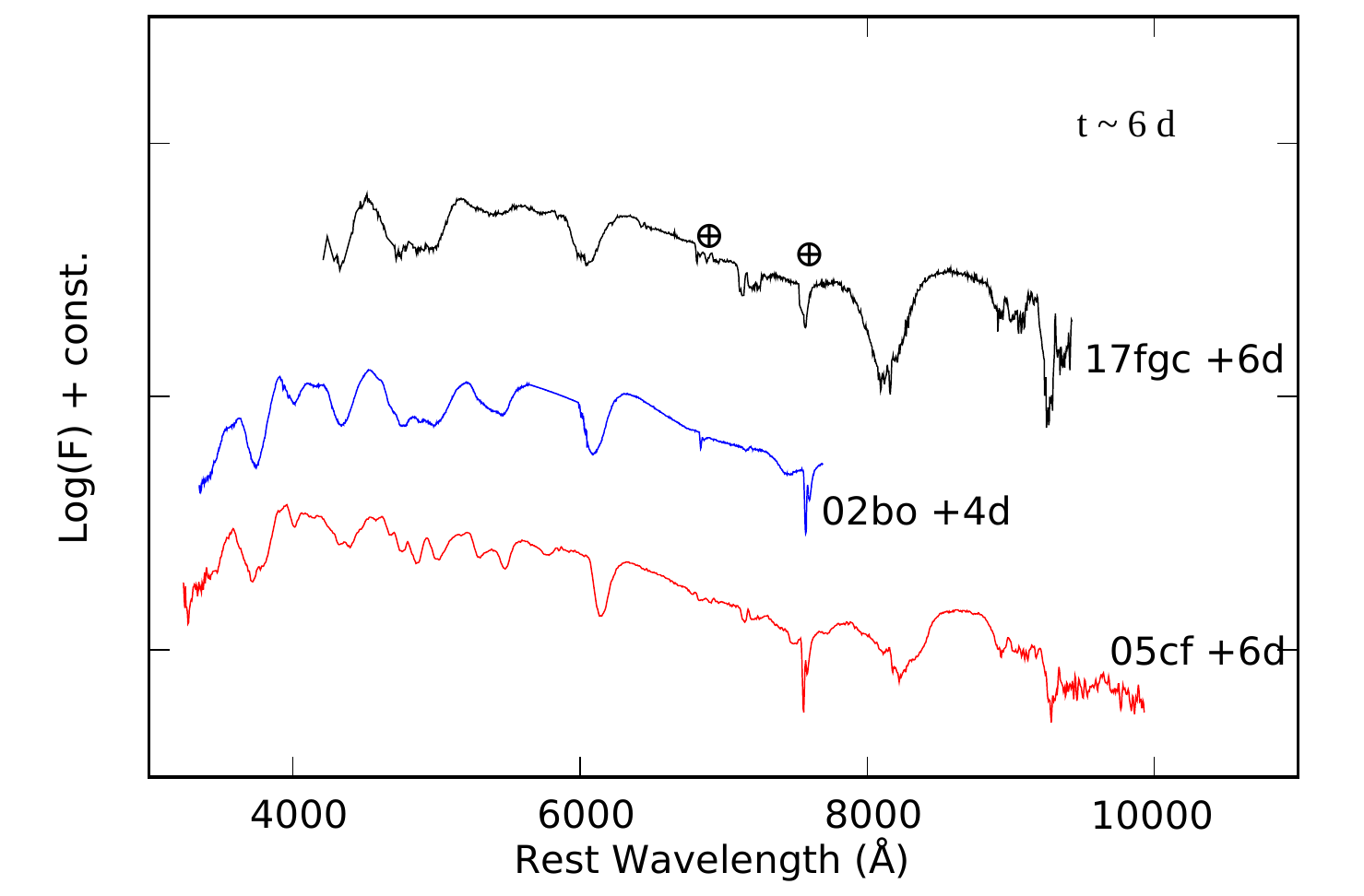}
   \end{subfigure}
   \begin{subfigure}[b]{0.4\linewidth}
     \includegraphics[width=\linewidth]{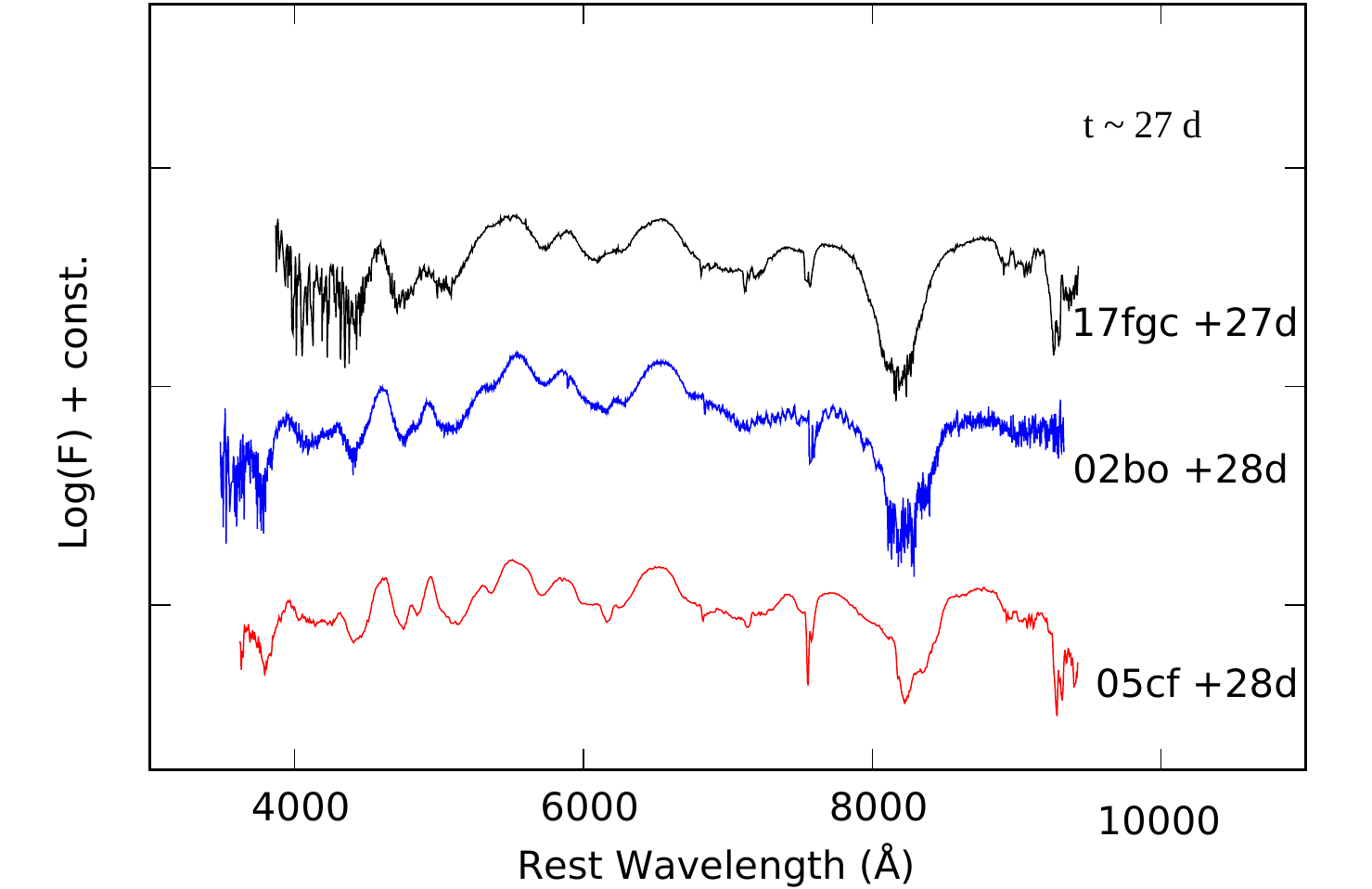}
   \end{subfigure}
   \begin{subfigure}[b]{0.4\linewidth}
     \includegraphics[width=\linewidth]{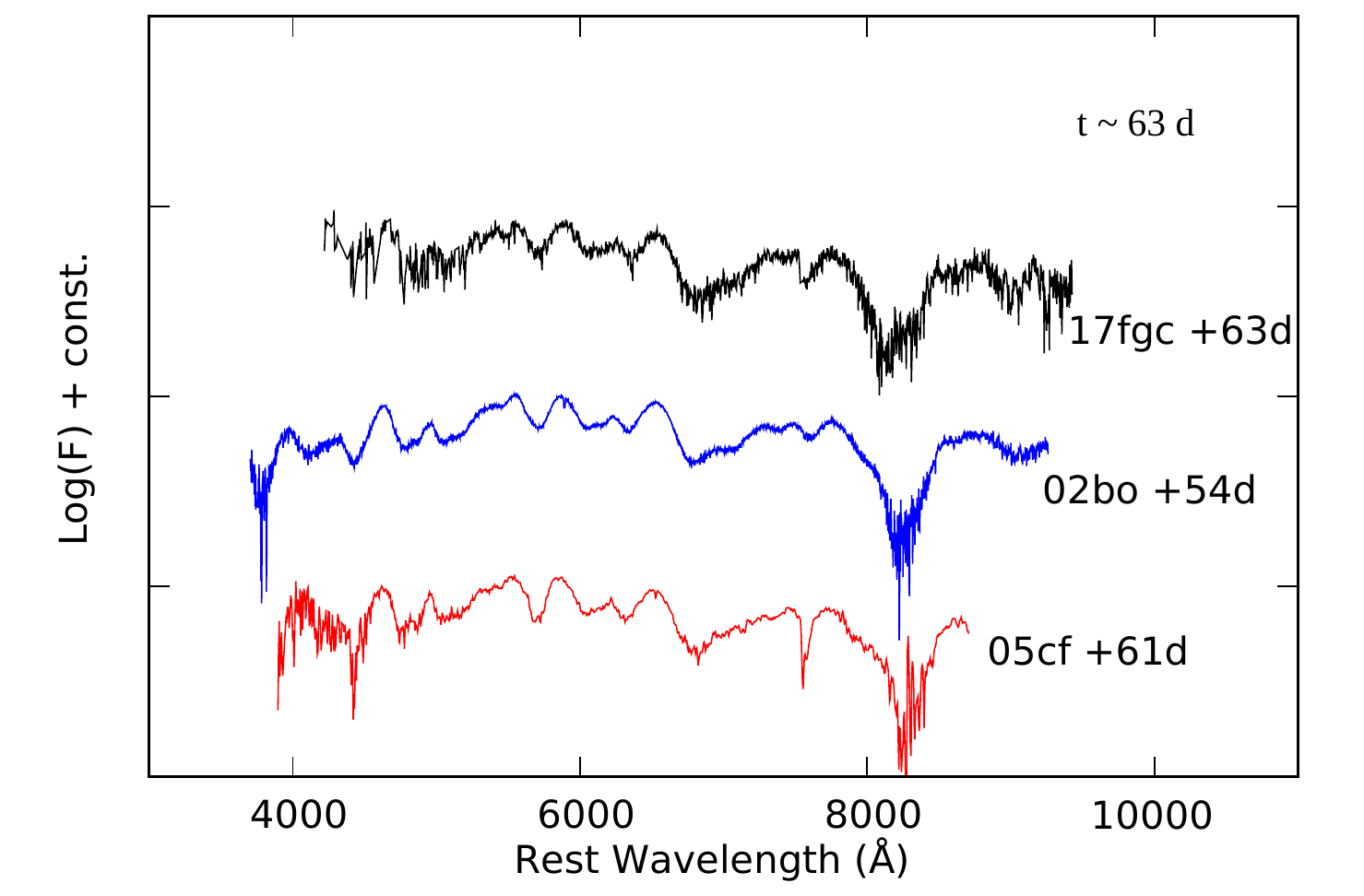}
   \end{subfigure}
   \caption{Spectra of SN 2017fgc at t $\simeq$ -5, +6, +27, +63 days after the $B$-band maximum overplotted with those of SN 2002bo and SN 2005cf at similar epochs (see the text for references). Telluric lines are showed with a mark.}
   \label{fig:SpecComp}
 \end{figure*}



\bsp	
\label{lastpage}
\end{document}